\documentclass[aps,prb,twocolumn,superscriptaddress,print]{revtex4-2}
\usepackage{amssymb,amsmath}
\usepackage{makecell}
\usepackage{rotating}
\usepackage{graphicx}
\usepackage{multirow}
\usepackage{hyperref}
\usepackage{appendix}
\usepackage{mathtools}
\usepackage{verbatim}
\usepackage{gensymb}
\usepackage{subfigure}
\usepackage{bm}
\usepackage{threeparttable}
\usepackage{booktabs}
\usepackage{braket}
\usepackage{xcolor}
\usepackage[normalem]{ulem}
\usepackage[mathscr]{euscript}
\usepackage{tabulary}
\usepackage{ragged2e}
\usepackage{adjustbox}
\usepackage{tabularx}
\usepackage{enumitem}
\usepackage{chngcntr}
\usepackage{capt-of}
\setlist{nolistsep}

\setlist{
  noitemsep,
  topsep=0pt,    
  partopsep=0pt,  
  parsep=0pt,     
  itemsep=0pt      
}

\begin{document}
\title{Symmetry-directed electronic and optical properties in a two-dimensional square-lattice ZnPc-MOF}
\author{Zhonghui Han}
\affiliation{Center for Quantum Sciences and School of Physics, Northeast Normal University, Changchun 130024, China}
\author{Lanting Feng}
\affiliation{College of Material Science and Engineering, Key Laboratory of Advanced Structural Materials, Ministry of Education, Changchun University of Technology, Changchun, 130012, China}
\author{Guodong Yu}
\email{yugd000@nenu.edu.cn}
\affiliation{Center for Quantum Sciences and School of Physics, Northeast Normal University, Changchun 130024, China}
\author{Shengjun Yuan}
\affiliation{Key Laboratory of Artificial Micro- and Nano-structures of Ministry of Education and School of Physics and Technology, Wuhan University, Wuhan 430072, China}
\affiliation{Wuhan Institute of Quantum Technology, Wuhan 430206, China}
\affiliation{School of Artificial Intelligence, Wuhan University, Wuhan 430072, China}

\begin{abstract}
The electronic structure of materials is fundamentally governed by their crystal symmetry. While most research on two-dimensional materials has focused on hexagonal lattices, such as graphene, hexagonal boron nitride, and transition metal dichalcogenides. This work explores a square-lattice system: the experimentally realized phthalocyanine-based metal–organic framework (ZnPc-MOF). Using group representation theory, we classify the electronic bands of ZnPc-MOF monolayer, AA- and AB-stacked bilayers, and twisted bilayers in terms of the irreducible representations (irreps) of their little groups. We find that bands in the AB-stacked bilayer remain two-fold degenerate along the $Y$ and $Y^{\prime}$ high-symmetry lines, as a consequence of the sole presence of two-dimensional irreps along these directions. We further derive optical transition selection rules to interpret the optical conductivity, revealing pronounced polarization-dependent optical responses. Additionally, we investigate the quasicrystalline electronic states in the 45$^{\circ}$ twisted bilayer (ZnPc-MOF quasicrystal) using the resonant coupling Hamiltonian. Compared to graphene quasicrystals, ZnPc-MOF quasicrystal exhibits weaker resonant coupling strengths, yet its quasicrystalline states lie closer to the Fermi energy, suggesting a greater contribution to low-energy electronic phenomena.
\end{abstract}

\maketitle
\section{Introduction}
The discovery of graphene marked the dawn of a new era in materials science, triggering extensive exploration into two-dimensional (2D) systems\cite{novoselov2004electric}. These materials exhibit extraordinary properties, such as atomic-scale thickness, strong quantum confinement, and unique electronic excitations, that are often absent in their bulk counterparts. Chief among these features is the electronic structure, which serves as the fundamental determinant of a material’s electronic, optical, and topological behaviors. It is well established that the electronic structure is intimately tied to the crystal symmetry, giving rise to the powerful concept of symmetry-property relationships in solid state physics. A well-known example is topological band theory, which connects the emergence of topological states to specific symmetries, including time-reversal, chiral, and crystalline symmetries\cite{hasan2010colloquium, qi2011topological}. More generally, essential features including energy-level degeneracy, orbital hybridization, and optical transition selection rules are directly governed by symmetry.

To date, the majority of well-studied 2D materials, such as graphene, hexagonal boron nitride (h-BN), and transition metal dichalcogenides (TMDs), crystallize in a hexagonal lattice\cite{geim2007rise,hBN_1stPaper,TMDC_first_paper}. In graphene, the coexistence of time-reversal, $C_3$ rotational, and inversion symmetries protects the massless Dirac fermions\cite{castro2009electronic}. Breaking inversion symmetry, as in h-BN, where inequivalent boron and nitrogen sites lift energy degeneracy, opens a band gap\cite{giovannetti2007substrate}. In TMDs, the hexagonal Brillouin zone (BZ) hosts degenerate but inequivalent $K$ and $K^{\prime}$ valleys, enabling phenomena such as spin-valley locking and valley-selective optical excitation\cite{xiao2012coupled}. These hallmark properties are thus deeply rooted in the symmetry character.

In contrast, square-lattice 2D materials remain comparatively less explored, largely due to their scarcity among naturally occurring or readily exfoliable crystals. Most studies have predominantly been confined to theoretical lattice models~\cite{altermagSquareLattice}, atom arrays~\cite{AtomArraysquarelattice}, or photonic crystals~\cite{PhotonicCrystalSquareLattice}. However, recent advances in synthetic chemistry, particularly in the realm of 2D metal–organic frameworks (MOFs), have enabled the fabrication of 2D materials with controlled lattice symmetries~\cite{2DMOF-symm0, 2DMOF1, 2D-MOF-conductive, 2DMOF-2007}. Composed of metal nodes coordinated by organic linkers, 2D MOFs form crystalline porous structures that offer high design flexibility and tunability. Importantly, 2D MOFs with various lattice geometries, including square, hexagonal, and Kagom\'{e} lattices, have been successfully synthesized~\cite{2DMOF-Kagome,2DMOF-symm0,2DMOF-qubit}. Consequently, 2D MOFs provide an ideal platform for investigating how the underlying lattice symmetry, particularly the less-explored square symmetry, dictates electronic and optical properties.

In this work, we present a comprehensive symmetry-based theoretical investigation of a prototypical square-lattice 2D MOF, specifically a zinc- and phthalocyanine-based framework (denoted ZnPc-MOF), which has been realized experimentally~\cite{MPc-MOF0, MPc-MOF1, MPc-MOF2, MPc-MOF3}. Employing symmetry analysis and group representation theory, we elucidate how the square-lattice symmetry governs the fundamental electronic structure and optical responses of this material. Beyond the monolayer, stacking configurations provide an additional degree of freedom for engineering symmetry and emergent properties. For instance, monolayer graphene exhibits point group $\mathbb{C}_{6v}$, whereas AA- and AB-stacked bilayers transform according to $\mathbb{D}_{6h}$ and $\mathbb{D}_{3d}$, respectively. A relative twist between layers can further reduce the symmetry, with profound consequences for interlayer hybridization~\cite{interlayer_rule_Yu} and correlated phenomena. The discovery of superconductivity and correlated insulating states in magic-angle twisted bilayer graphene ($\sim 1.1^\circ$)~\cite{supercond_tBG, correlated_semicond_tBG}, associated with the emergence of flat bands~\cite{flatband_tBG}, has catalyzed extensive research into twisted and Moir\'{e} systems\cite{tB_Bravais_lattices,tB_BN,tB_MoS2,tB_TMD,tB_gy,gyu_njp_2024_1}. Furthermore, a twist angle of $30^\circ$ renders bilayer graphene incommensurate, giving rise to a quasicrystal with point group $\mathbb{D}_{6d}$~\cite{gqc_science, gqc_pnas, gqc_npj}. This concept of van der Waals (vdW) quasicrystals has subsequently been extended to other layered systems recently, including TMDs~\cite{tmd_qc_NP, tmd_qc_nature} and 2D MOFs~\cite{KimMOFQC2025}. Given that different stacking configurations (e.g., AA-stacked, AB-stacked, and twisted) of ZnPc-MOF bilayers modify the overall symmetry, we extend our analysis to encompass various bilayer structures, with particular attention to the $45^\circ$ twisted bilayer, which forms a vdW quasicrystal exhibiting eightfold rotational symmetry.

Recently, altermagnetism in 2D square-lattice materials has attracted considerable attention~\cite{altermag_0, altermag_1, altermag_2, altermag_3, altermag_4, altermag_5}. In addition, there are also studies examining other aspects of non-hexagonal 2D lattices, such as the symmetry-indicator analysis for filling anomalies in $C_4$-symmetric square lattices~\cite{C4_filling_anomaly}, the realization of Hubbard models in Moir\'{e} 2D square lattices~\cite{moire_square_latt}, numerical investigations of the electronic structure of S-graphene (which features a rectangular lattice)~\cite{S-graphene}, and the construction of symmetry-based model Hamiltonians for the topological characterization of square-lattice systems~\cite{model_hamil_square_latt}, among others. Despite these efforts, a systematic symmetry analysis of the fundamental band structure and optical selection rules for a realistic 2D square-lattice material has remained absent. The present work provides such an analysis, using ZnPc-MOF as a concrete and experimentally relevant example. More broadly, we establish a general symmetry-based theoretical framework for understanding square-lattice 2D materials. The conclusions drawn are applicable to any material sharing the same space group symmetry, offering a predictive tool for exploring this emerging family of materials.

The remainder of this paper is organized as follows. In Section II, we introduce the tight-binding model for ZnPc-MOF-based systems. Section III.A presents the symmetry analysis and band structure classification for the monolayer and bilayers. We derive the optical transition selection rules (OTSRs) and explain the optical conductivity using the OTSR in Section III.B. Section IV is devoted to the quasicrystalline electronic states in the ZnPc-MOF quasicrystal. We conclude in Section V with a summary of our findings.

\begin{figure}[!htbp]
\centering
\includegraphics[width=8.5 cm]{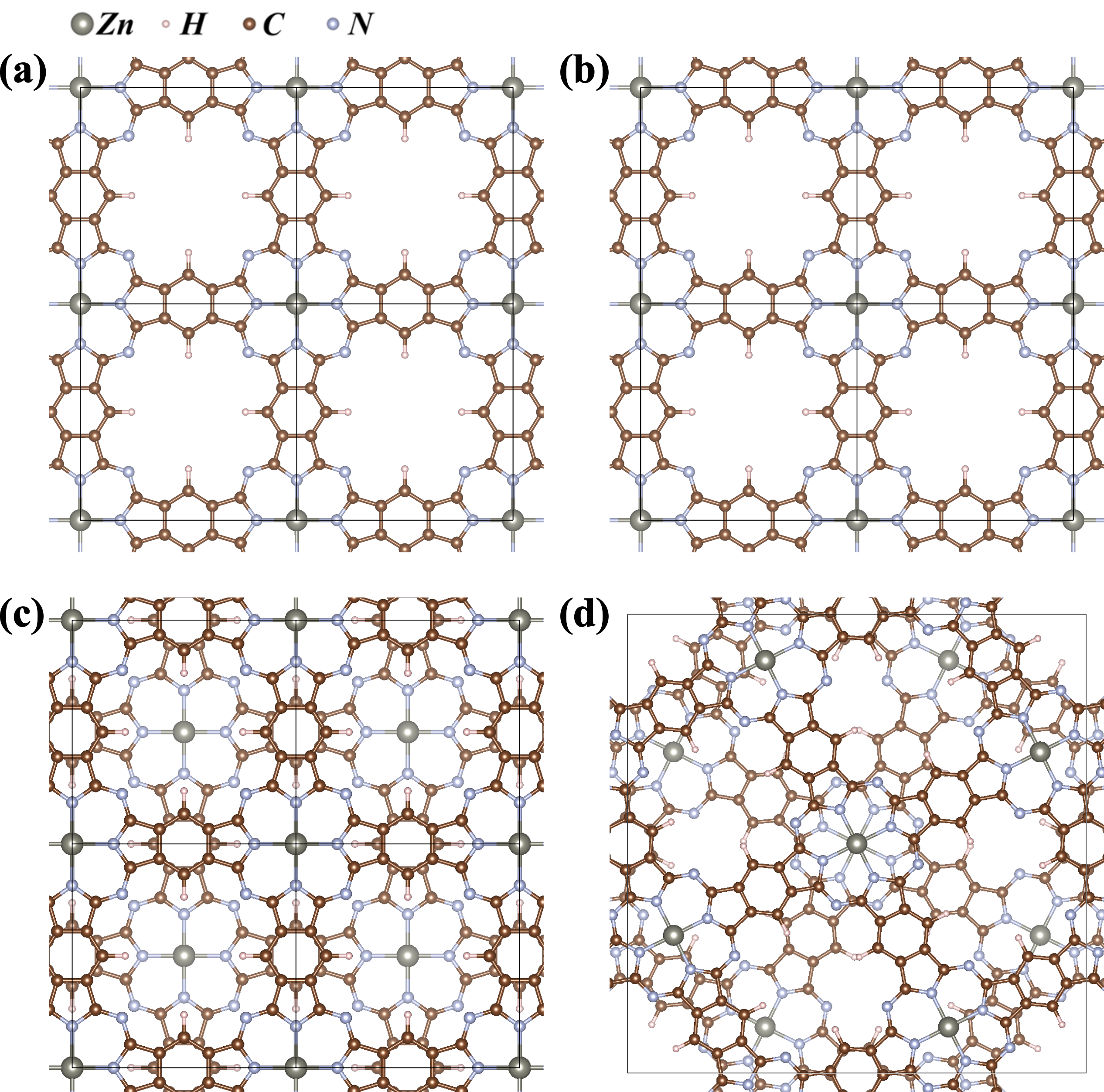}
\caption{Structures of ZnPc-MOF (a) monolayer, (b)AA-stacked, (c) AB-stacked and (d) $36.87^{\circ}$ twisted bilayer $(m=2,n=1)$. Four unit cells are shown in (a)-(c), and one unit cell in (d). The middle point of each subfigure (Zn atom position) is as the origin when we construct the symmetry operation matrixes.}
\label{fig:struct}
\end{figure}

\begin{figure*}[!htbp]
\centering
\includegraphics[width=0.8\textwidth]{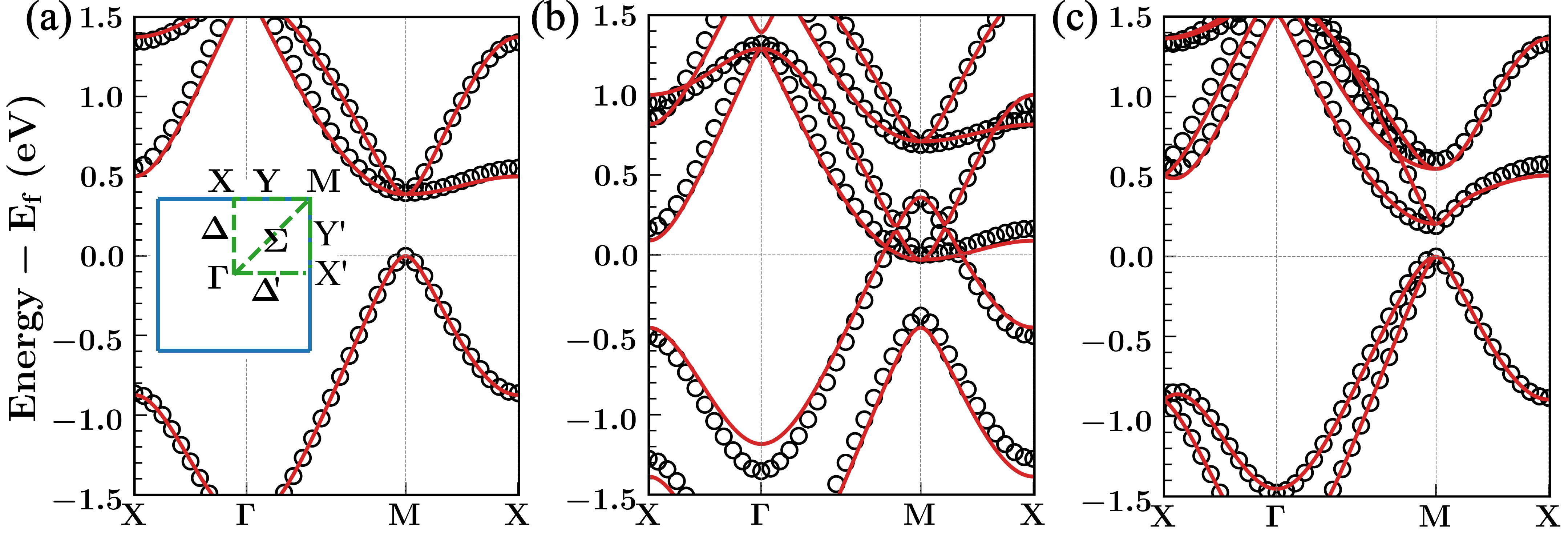}
\caption{Band structures of (a) monolayer, (b) AA-stacked, and (c) AB-stacked bilayer. Empty circles and solid red lines represent band structures obtained from DFT calculations and tight-binding model, respectively. The vdW correction is included in all DFT calculations for bilayer structures. The inset in (a) displays the BZ and high-symmetry points and lines.}
\label{fig:band_dft}
\end{figure*}

\section{Structures and tight-binding model}
\textit{Structures.---} The ZnPc-MOF monolayer exhibits a square lattice structure, as illustrated in Fig.~\ref{fig:struct}(a). The lattice vectors are given by $\bm{a}_1 = a\bm{i}$ and $\bm{a}_2 = a\bm{j}$, where the lattice constant $a = 10.75~{\AA}$. A bilayer is formed by stacking one layer on top of another with various configurations such as AA-, AB-, or twisted stacking, shown in Fig.~\ref{fig:struct}(b)--(d). The interlayer distance is set to 3.38 {\AA} for all bilayer structures through this paper unless stated otherwise. For commensurate twisted bilayers that preserve translational symmetry, the twist angle $\theta$ is defined by two integers $m$ and $n$ through the relation:
\[
\theta = \arccos \left( \frac{2mn}{m^2 + n^2} \right).
\]
The twist angle is measured relative to the AA stacking configuration. The corresponding lattice vectors of the twisted bilayer then become:
\[
\bm{a}'_{1} = n\bm{a}_1 + m\bm{a}_2, \quad \bm{a}'_{2} = -m\bm{a}_1 + n\bm{a}_2.
\]

\textit{Symmetries.---} In this work, symmetry operations are labeled according to the following conventions: $E$ denotes the identity operation; $i$ represents inversion; $C_{n\alpha}^m$ indicates a rotation by $2\pi m/n$ anticlockwise about an axis along direction $\alpha$, where $\alpha \in \{x, y, z, xy, \overline{x}y\}$ corresponds to the directions $\bm{i}$, $\bm{j}$, $\bm{k}$, $\bm{i}+\bm{j}$, and $-\bm{i}+\bm{j}$, respectively; $\sigma_{\alpha\beta}$ refers to a mirror reflection with respect to a plane spanned by the two vectors specified by $\alpha$ and $\beta$—for instance, $xy$ denotes the mirror plane containing $\bm{i}$ and $\bm{j}$, while $(xy)z$ denotes the plane containing $\bm{i}+\bm{j}$ and $\bm{k}$; and $S_{nz}^{m} = \sigma_{xy} C_{nz}^m$ defines an improper rotation.

The monolayer, AA-stacked bilayer, and twisted bilayer structures have symmorphic space groups (SSGs) $P4mm$, $P4/mmm$, and $P422$, respectively. Their symmetry operations can be described by the semi-direct product $\mathbb{T} \rtimes \mathbb{G}_0$ of the translation group $\mathbb{T}$ and the point group $\mathbb{G}_0$, where $\mathbb{G}_0$ corresponds to $\mathbb{C}_{4v}$ for the monolayer, $\mathbb{D}_{4h}$ for the AA-stacked bilayer, and $\mathbb{D}_4$ for the twisted bilayer. In contrast, the AB-stacked bilayer possesses a non-symmorphic space group (non-SSG), $P4/nmm$. Its symmetry operations decompose into right cosets of the translation group as $\sum_i \mathbb{T} g_i$, and the elements $g_i$ constitute the set $
\mathbb{G}_0 = \mathbb{C}_{4v} + \{\sigma_{xy}|\bm{\tau}_0\} \mathbb{C}_{4v}$, where $\{\sigma_{xy}|\bm{\tau}_0\}$ represents a mirror reflection across the $xy$-plane followed by a translation $\bm{\tau}_0 = \frac{1}{2}\bm{a}_1 + \frac{1}{2}\bm{a}_2$. Note that due to the non-zero translation $\bm{\tau}_0$, the set $\mathbb{G}_0$ for AB-stacked bilayer does not form a group.

\begin{figure*}[!htbp]
\centering
\includegraphics[width=1.\textwidth]{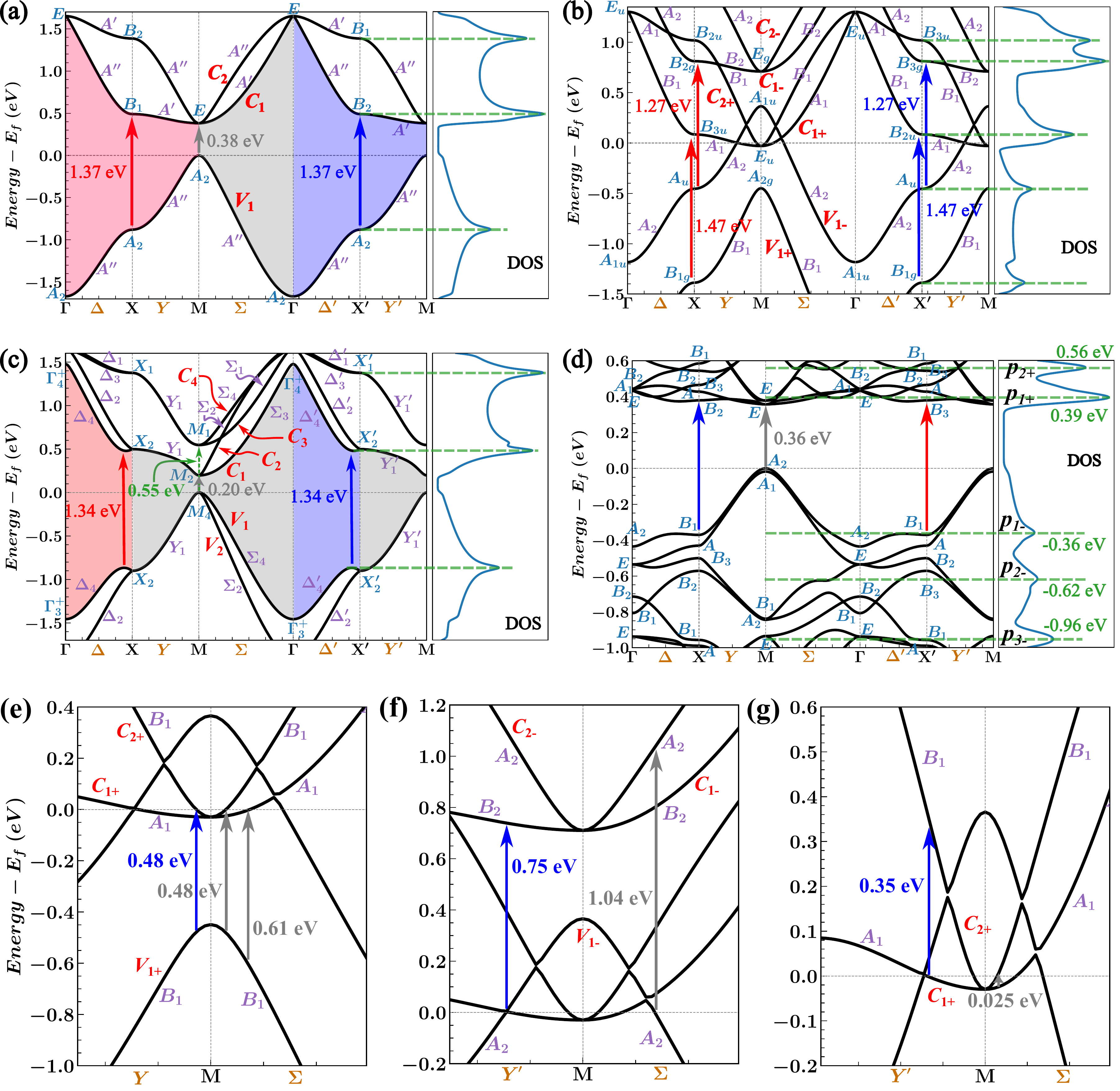}
\caption{Irreps of electronic states at high-symmetry points and along high-symmetry lines for (a) monolayer, (b) AA-stacked, (c) AB-stacked, and (d) $36.78^\circ$ twisted bilayer. Irreps along high-symmetry lines are omitted in (d) due to the high density of closely spaced bands. The three bands of the monolayer in (a) are labeled $V_1$, $C_1$, and $C_2$. The six bands of the AA-stacked bilayer in (b) are labeled $V_{1\pm}$, $C_{1\pm}$, and $C_{2\pm}$, where $\pm$ denotes bonding/anti-bonding character. The six bands of the AB-stacked bilayer in (c) are labeled $V_1$, $V_2$, $C_1$, $C_2$, $C_3$, and $C_4$. In (a) and (c), shaded regions highlight $V_1 \rightarrow C_1$ interband transitions, with blue, red, and gray indicating transitions allowed by $x$-polarized, $y$-polarized, and both polarized lights, respectively. (e) and (f) show minimum transition energies for interband transitions $V_{1+} \rightarrow C_{i+}$ and $V_{1-} \rightarrow C_{i-}$ ($i = 1,2$) in the AA-stacked bilayer. (g) shows minimum and maximum transition energies for interband transition $C_{1+} \rightarrow C_{2+}$ in the AA-stacked bilayer. Blue, red, and gray arrows mark transitions excited by $x$-polarized, $y$-polarized, and both polarizations, respectively.}
\label{fig:band_irrep}
\end{figure*}

\textit{Tight-binding model.---} For the ZnPc-MOF monolayer, the electronic states near the Fermi energy are dominated solely by the $p_z$ orbitals of the carbon and nitrogen atoms\cite{MPc-MOF3}. This allows us to construct a tight-binding model based on $p_z$ orbitals to describe the low-energy electronic states for both monolayer and bilayers with arbitrary stacking configurations. The Hamiltonian is given by:
\begin{equation}
\label{eq:H}
H = \sum_{i \in \{\mathrm{C}\}} \varepsilon_C C_i^{\dagger} C_i + \sum_{i \in \{\mathrm{N}\}} \varepsilon_N C_{i}^{\dagger} C_{i} + \sum_{i \neq j} t(\bm{r}_{ij}) C_i^{\dagger} C_j,
\end{equation}
where $C_i^{\dagger}$ and $C_i$ are the creation and annihilation operators for the $p_z$ orbital at site $i$, $\varepsilon_C$ and $\varepsilon_N$ are the on-site energies for carbon and nitrogen sites, respectively, and $t(\bm{r}_{ij})$ is the hopping integral between $p_z$ orbitals at sites $i$ and $j$, with $\bm{r}_{ij}$ being the relative position vector. The hopping integral can be decomposed into contributions from $\pi$ and $\sigma$ bonds as~\cite{LCAO}:
\begin{equation}
t(\bm{r}_{ij}) = n^2 V_{pp\sigma}(|\bm{r}_{ij}|) + (1 - n^2) V_{pp\pi}(|\bm{r}_{ij}|),
\end{equation}
where $n$ is the direction cosine of $\bm{r}_{ij}$ with respect to the $z$-axis. The Slater--Koster parameters $V_{pp\sigma}$ and $V_{pp\pi}$ are given by:
\begin{equation}
\begin{split}
V_{pp\pi}(|\bm{r}|) &= -\gamma_0 e^{\delta(a_0 - |\bm{r}|)}, \\
V_{pp\sigma}(|\bm{r}|) &= \gamma_1 e^{\delta(h_0 - |\bm{r}|)}.
\end{split}
\end{equation}
In this work, we employ the following parameter set: $\varepsilon_C = 0.0$~eV, $\varepsilon_N = -2.2$~eV, $\gamma_0 = 3.4$~eV, $\gamma_1 = 0.6$~eV, $a_0 = 1.42$~\AA, $h_0 = 3.35$~\AA, and $\delta = 2.4$~\AA$^{-1}$. All hopping terms with an interatomic distance larger than $5$~{\AA} are neglected. As demonstrated in Fig.~\ref{fig:band_dft}, this tight-binding model successfully reproduces the band structures of both monolayer and bilayer ZnPc-MOF obtained from density functional theory (DFT) calculations~\cite{dft0,dft1} implemented in VASP software\cite{vasp}, using the HSE06 hybrid functional~\cite{hse06} with vdW corrections~\cite{vdw0} included. 

\section{Electronic and optical properties}
\subsection{Band structures and band labeling}
Diagonalizing the Hamiltonian at $\bm{k}$ constructed as Appendix \ref{app:Hk}, we can obtain the band structures of the monolayer, AA-stacked, AB-stacked and twisted bilayer. The results reveal that the monolayer is a semiconductor with a band gap of $0.38$~eV. AA stacking induces a semiconductor-to-semimetal transition with a negative band gap -0.4 eV, which will be explained in detail later, while the AB-stacked bilayer, in contrast, remains semiconducting but with a reduced band gap of $0.2$~eV. Twisted bilayer with large twist angle almost keep the band gap of monolayer. The states can be labeled by the irreducible representations (irreps) of the little groups according to Appendix \ref{app:irrep}. The character tables with the irrep definition of all systems under consideration in this paper are listed in tables of Appendix \ref{app:tables}.

Figure~\ref{fig:band_irrep} displays the band structures, with each band labeled by the corresponding irrep of the little group. For monolayer, AA-stacked, and twisted bilayer with SSG symmetry, the little group is isomorphic to a crystallographic point group, and we accordingly label the irreps using standard point group notation. For the AB-stacked bilayer with non-SSG symmetry, the irreps are labeled following the convention of the Bilbao Crystallographic Server~\cite{BCS_nonSSP_ir}. These irrep assignments are consistent with the compatibility relationships (CRs) between each symmetry line and its endpoints, as detailed in Table~\ref{tab:compi_relationship}.

\subsubsection{Band structure change from monolayer to AA- and AB-stacked bilayer.}
The evolution of the band structure from a monolayer to AA- and AB-stacked bilayers can be understood in terms of interlayer hybridization and symmetry. To facilitate the discussion, we denote the highest valence band (VB) of the monolayer as $V_1$, and the two lowest conduction bands (CBs) as $C_1$ and $C_2$, as illustrated in Fig.~\ref{fig:band_irrep}(a). In the absence of interlayer coupling, the two layers exhibit degenerate energy spectra. The interlayer coupling introduces hybridization between the VBs of the two layers (interlayer VB–VB hybridization) and between their CBs (interlayer CB–CB hybridization). In contrast, hybridization between the VB of one layer and the CB of the other (interlayer VB–CB hybridization) is negligible, owing to the energy separation imposed by the band gap. Therefore, the interlayer coupling can be effectively described as separate interlayer VB–VB and CB–CB hybridizations.\cite{gyu_prb_2023,gyu_njp_2024_1}

\textit{AA-stacked bilayer.---} From monolayer to AA-stacked bilayer, a configuration with the strongest interlayer coupling due to the direct vertical overlap of two layers' $p_z$ orbitals, the interlayer VB-VB and CB-CB hybridizations lift the degeneracy of two layers, splitting both the VBs and CBs into bonding states (shifted downward in energy) and anti-bonding states (shifted upward). As shown in Fig. \ref{fig:band_irrep}(b), the $V_1$ band of monolayer split to the one bonding band ($V_{1+}$ band) and one anti-bonding band ($V_{1-}$ band), and $C_1$ band split to one $C_{1+}$ band and one $C_{1-}$ band, and $C_2$ band split to one $C_{2+}$ band and one $C_{2-}$ band. Consequently, the density of state (DOS) peaks near the Fermi energy in both the electron and hole regions of the monolayer split into two corresponding peaks in the bilayer. At the $M$ point, the valence band maximum (VBM) of monolayer with irrep $A_2$ splits into bonding and anti-bonding state with irreps $A_{2g}$ and $A_{1u}$, respectively. Similarly, the conduction band minimum (CBM), originally corresponding to the irrep $E$, splits into bonding and anti-bonding states with irreps $E_u$ and $E_g$, respectively. The semiconductor-to-semimetal transition from the monolayer to the AA-stacked bilayer is driven by anti-bonding $A_{1u}$ VB state rising above the bonding $E_u$ CB state. This phenomenon can be described by the negative band gap, which is defined by the value $\delta \varepsilon=E(E_u)-E(A_{1u})$. Similar to the semiconducting AB-stacked and twisted bilayer, smaller $\delta \varepsilon$ means a stronger interlayer coupling strength.

\textit{AB-stacked bilayer.---} Compared to the AA-stacked bilayer, the interlayer coupling in the AB-stacked configuration is weaker due to the reduced overlap between the $p_z$ orbitals of the two layers. Consequently, the band splitting induced by interlayer coupling is less pronounced in the AB-stacked bilayer than in its AA-stacked counterpart. The main features of the band structure evolution from the monolayer can be understood through symmetry considerations, and the hybridization process can be described as follows. In the absence of interlayer coupling, the two layers possess identical band structures. Upon coupling, each band along the high-symmetry lines splits into two closely spaced bands. However, no splitting occurs along the $Y$ and $Y^{\prime}$ lines and at the $X$, $X^{\prime}$, and $M$ points. This degeneracy is protected by the existence of only 2D irreps at these symmetry lines and points. An additional correction is required at the $M$ point: the fourfold degenerate energy level at $M$ (comprising two $E$ levels in the absence of interlayer coupling) splits into two twofold degenerate levels (one $M_1$ and one $M_2$ level), as only 2D irreps are available at the $M$ point for the AB-stacked bilayer. This splitting reduces the band gap from 0.38 eV to 0.2 eV. As a result, the two $V_1$ bands from the individual monolayers evolve into one $V_1$ and one $V_2$ band in the AB-stacked bilayer, while the two $C_1$ and two $C_2$ bands hybridize to form the $C_1$, $C_2$, $C_3$, and $C_4$ bands shown in Fig.~\ref{fig:band_irrep}(c). The twofold degeneracy observed along the $Y$ and $Y^{\prime}$ lines and at the $X$, $X^{\prime}$, and $M$ points is a distinctive feature of square-lattice systems and contrasts with hexagonal bilayers such as graphene. In both AA- and AB-stacked bilayer graphene, which preserve the SSG symmetry, the little group at any wavevector always contains one-dimensional (1D) irreps.

\subsubsection{Irrep relations between symmetrically equivalent wavevectors}
Owing to symmetry, the Hamiltonians at a wavevector $\bm{k}$ and its symmetrically equivalent wavevector $\bm{k}^{\prime} = \sigma_{(xy)z}\bm{k}$ share identical energy spectra. However, the energy levels at $\bm{k}$ and $\bm{k}^{\prime}$ are not always assigned the same irrep. By comparing the high-symmetry points and lines $X$, $Y$, $\Delta$ with their symmetrically equivalent counterparts $X^{\prime}$, $Y^{\prime}$, $\Delta^{\prime}$, one finds that the energy spectra at $\bm{k}$ ($=Y,\Delta$) and $\bm{k}^{\prime}$ ($=Y^{\prime},\Delta^{\prime}$) exhibit the same irreps. In contrast, for $X$ and $X^{\prime}$, different irreps may be assigned to certain degenerate energy levels. We denote this irrep transformation as
\begin{equation}
\label{eq:ir1_ir2_X_Xp}
\begin{split}
&B_{1} \leftrightharpoons B_{2} \quad \text{(Monolayer)}, \\
&A_{2g} \leftrightharpoons A_{3g},\quad A_{2u} \leftrightharpoons A_{3u} \quad \text{(AA-bilayer)},\\
&B_{2} \leftrightharpoons B_{3} \quad \text{(Twisted bilayer)},
\end{split}
\end{equation}
where $ir_1 \leftrightharpoons ir_2$ indicates that, under the $\sigma_{(xy)z}$ operation, the energy levels with irrep $ir_1$ at $X$ and $X^{\prime}$ transform into levels with irrep $ir_2$ at $X^{\prime}$ and $X$, respectively. This irrep change can be derived analytically using symmetry operations, as demonstrated below for the monolayer case. We use $|X, ir\rangle$ to denote a state at $X$ with irrep $ir$, and $|X^{\prime}, ir^{\prime}\rangle$ its degenerate partner at $X^{\prime}$ with irrep $ir^{\prime}$. These are related by $|X^{\prime}, ir^{\prime}\rangle = \sigma_{(xy)z} |X, ir\rangle$. From Table~\ref{tab:irrep_monolayer}, the symmetry operations act on the states at $X$ as:
\begin{equation}
\begin{split}
(E, C_{2z}, \sigma_{xz}, \sigma_{yz})|X,A_1\rangle &= (1,1,1,1)|X,A_1\rangle, \\
(E, C_{2z}, \sigma_{xz}, \sigma_{yz})|X,A_2\rangle &= (1,1,-1,-1)|X,A_2\rangle, \\
(E, C_{2z}, \sigma_{xz}, \sigma_{yz})|X,B_1\rangle &= (1,-1,1,-1)|X,B_1\rangle, \\
(E, C_{2z}, \sigma_{xz}, \sigma_{yz})|X,B_2\rangle &= (1,-1,-1,1)|X,B_2\rangle.
\end{split}
\end{equation}
Furthermore, the following commutation relations hold:
\begin{equation}
\begin{split}
\sigma_{(xy)z} E &= E \sigma_{(xy)z}, \\
\sigma_{(xy)z} C_{2z} &= C_{2z} \sigma_{(xy)z}, \\
\sigma_{(xy)z} \sigma_{xz} &= \sigma_{yz} \sigma_{(xy)z}, \\ 
\sigma_{(xy)z} \sigma_{yz} &= \sigma_{xz} \sigma_{(xy)z}.
\end{split}
\end{equation}
Consequently, the symmetry operators act on the rotated states as:
\begin{equation}
\begin{split}
(E, C_{2z}, \sigma_{xz}, \sigma_{yz}) \sigma_{(xy)z}|X,A_1\rangle &= (1,1,1,1) \sigma_{(xy)z}|X,A_1\rangle, \\
(E, C_{2z}, \sigma_{xz}, \sigma_{yz}) \sigma_{(xy)z}|X,A_2\rangle &= (1,1,-1,-1) \sigma_{(xy)z}|X,A_2\rangle, \\
(E, C_{2z}, \sigma_{xz}, \sigma_{yz}) \sigma_{(xy)z}|X,B_1\rangle &= (1,-1,-1,1) \sigma_{(xy)z}|X,B_1\rangle, \\
(E, C_{2z}, \sigma_{xz}, \sigma_{yz}) \sigma_{(xy)z}|X,B_2\rangle &= (1,-1,1,-1) \sigma_{(xy)z}|X,B_2\rangle.
\end{split}
\end{equation}
This leads directly to the irrep correspondence:
\begin{equation}
\begin{split}
\sigma_{(xy)z}|X,A_1\rangle &= |X^{\prime},A_1\rangle, \\
\sigma_{(xy)z}|X,A_2\rangle &= |X^{\prime},A_2\rangle, \\
\sigma_{(xy)z}|X,B_1\rangle &= |X^{\prime},B_2\rangle, \\
\sigma_{(xy)z}|X,B_2\rangle &= |X^{\prime},B_1\rangle.
\end{split}
\end{equation}
The corresponding irrep change from $X^{\prime}$ to $X$ and for other systems can be demonstrated using an analogous approach and is omitted here for brevity. This transformation in irrep has significant implications for the OTSRs, as discussed in the following section.

\begin{figure}[!htbp]
\centering
\includegraphics[width=8.5 cm]{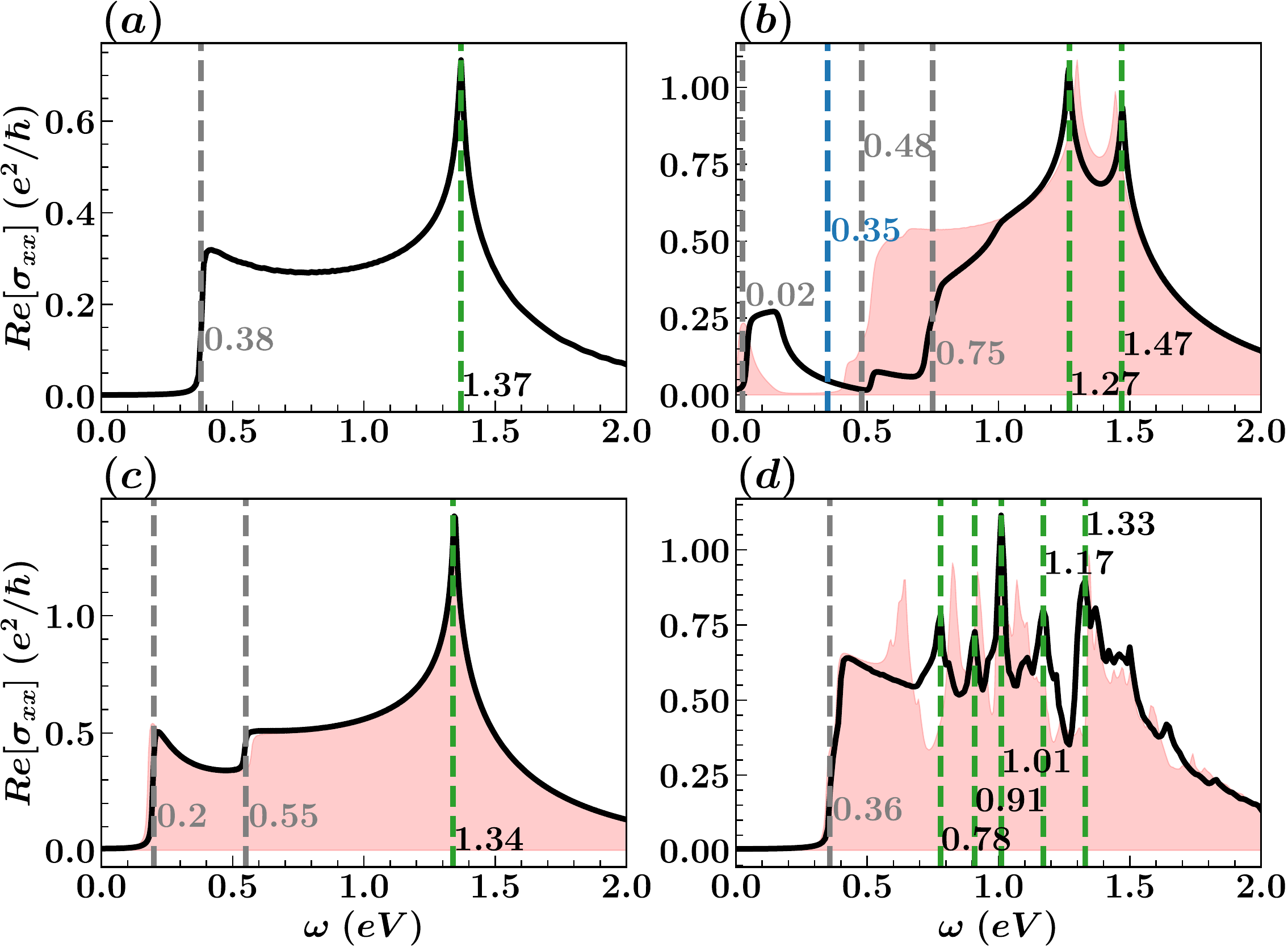}
\caption{Optical absorption spectra of charge neutral (a) monolayer, (b) AA-stacked, (c) AB-stacked, and (d) 36.87$^{\circ}$ twisted bilayer. The areas in shade for (b), (c) and (d) show the optical absorption spectra considering the relaxation. The vertical gray and green dashed lines show the starting energies of absorption steps and absorption peaks. The vertical blue dashed line in (b) shows the maximum transition energy for interband transition $C_{1+} \rightarrow C_{2+}$ as shown in Fig. \ref{fig:band_irrep} (g).}
\label{fig:ac}
\end{figure}

\begin{table*}[htb]
\centering
\caption{The OTSRs for monolayer, AA-stacked, AB-stacked, and twisted bilayer structures at high-symmetry points and lines under both $x$- and $y$-polarized light. The notation $ir \leftrightarrow ir^{\prime}$ indicates that an electron in the $ir$ state can be excited to the $ir^{\prime}$ state by $x$- or $y$-polarized light, and vice versa.}
\begin{tabular*}{\linewidth}{@{\extracolsep{\fill}}l|c|c@{}}
\toprule
  Monolayer      & $x-polarization$ & $y-polarization$  \\
\midrule
$\Gamma,M$ & $A_1,A_2,B_1,B_2 \leftrightarrow E$ & $A_1,A_2,B_1,B_2 \leftrightarrow E$ \\
$X,X^{\prime}$ & $A_1 \leftrightarrow B_1; A_2 \leftrightarrow B_2$ & $A_1 \leftrightarrow B_2; A_2 \leftrightarrow B_1$ \\
$Y,\Delta^{\prime}$ & $A^{\prime} \leftrightarrow A^{\prime}; A^{\prime\prime} \leftrightarrow A^{\prime\prime}$ & $A^{\prime} \leftrightarrow A^{\prime\prime}$ \\
$\Delta,Y^{\prime}$ & $A^{\prime} \leftrightarrow A^{\prime\prime}$& $A^{\prime} \leftrightarrow A^{\prime}; A^{\prime\prime} \leftrightarrow A^{\prime\prime}$ \\
$\Sigma$ & $A^{\prime},A^{\prime\prime} \leftrightarrow A^{\prime},A^{\prime\prime}$& $A^{\prime},A^{\prime\prime} \leftrightarrow A^{\prime},A^{\prime\prime}$ \\
\bottomrule
  AA-bilayer      & $x-polarization$ & $y-polarization$  \\
\midrule
$\Gamma,M$ & $A_{1g},A_{2g}, B_{1g}, B_{2g}\leftrightarrow E_u$; $A_{1u},A_{2u}, B_{1u}, B_{2u}\leftrightarrow E_g$ & $A_{1g},A_{2g}, B_{1g}, B_{2g}\leftrightarrow E_u$; $A_{1u},A_{2u}, B_{1u}, B_{2u}\leftrightarrow E_g$  \\
$X,X^{\prime}$ & $A_g \leftrightarrow B_{3u};B_{1g}\leftrightarrow B_{2u}$; $B_{2g} \leftrightarrow B_{1u};B_{3g}\leftrightarrow A_{u}$ & $A_g \leftrightarrow B_{2u};B_{1g}\leftrightarrow B_{3u}$;$B_{2g} \leftrightarrow A_{u};B_{3g}\leftrightarrow B_{1u}$ \\
$Y,\Delta^{\prime}$ &$A_1 \leftrightarrow A_1;A_2\leftrightarrow A_2$; $B_1 \leftrightarrow B_1;B_2 \leftrightarrow B_2$   & $A_1 \leftrightarrow B_1;A_2\leftrightarrow B_2$  \\
$\Delta,Y^{\prime}$ &$A_1 \leftrightarrow B_1;A_2\leftrightarrow B_2$ & $A_1 \leftrightarrow A_1;A_2\leftrightarrow A_2$; $B_1 \leftrightarrow B_1;B_2 \leftrightarrow B_2$  \\
$\Sigma$ & $A_1, B_1 \leftrightarrow A_1,B_1$; $A_2, B_2\leftrightarrow A_2,B_2$ & $A_1, B_1 \leftrightarrow A_1,B_1$; $A_2, B_2\leftrightarrow A_2,B_2$  \\
\bottomrule
  AB-bilayer      & $x-polarization$ & $y-polarization$  \\
\midrule
$\Gamma$ & $\Gamma_{1-4}^- \leftrightarrow \Gamma_5^+;\Gamma_{1-4}^+ \leftrightarrow \Gamma_5^-$ & $\Gamma_{1-4}^- \leftrightarrow \Gamma_5^+;\Gamma_{1-4}^+ \leftrightarrow \Gamma_5^-$ \\
$M$ & $M_1, M_2 \leftrightarrow M_3,M_4$ & $M_1, M_2 \leftrightarrow M_3,M_4$ \\
$X$ & $X_1 \leftrightarrow X_2$ & $X_1 \leftrightarrow X_1;X_2 \leftrightarrow X_2$   \\
$X^{\prime}$ & $X_1^{\prime} \leftrightarrow X_1^{\prime};X_2^{\prime} \leftrightarrow X_2^{\prime}$ & $X_1^{\prime} \leftrightarrow X_2^{\prime}$   \\
$Y^{(\prime)}$ &  $Y_1^{(\prime)} \leftrightarrow Y_1^{(\prime)}$ & $Y_1^{(\prime)} \leftrightarrow Y_1^{(\prime)}$   \\
$\Delta$ &  $\Delta_1 \leftrightarrow \Delta_2;\Delta_3 \leftrightarrow \Delta_4$ &$\Delta_i \leftrightarrow \Delta_i(i=1-4)$    \\
$\Delta^{\prime}$ &  $\Delta_i^{\prime} \leftrightarrow \Delta_i^{\prime}(i=1-4)$ &$\Delta_1^{\prime} \leftrightarrow \Delta_2^{\prime};\Delta_3^{\prime} \leftrightarrow \Delta_4^{\prime}$    \\
$\Sigma$ & $\Sigma_1,\Sigma_2 \leftrightarrow \Sigma_1,\Sigma_2$; $\Sigma_3,\Sigma_4 \leftrightarrow \Sigma_3,\Sigma_4$ &  $\Sigma_1,\Sigma_2 \leftrightarrow \Sigma_1,\Sigma_2$; $\Sigma_3,\Sigma_4 \leftrightarrow \Sigma_3,\Sigma_4$  \\
\bottomrule
  twisted bilayer      & $x-polarization$ & $y-polarization$  \\
\midrule
$\Gamma,M$ & $A_1,A_2,B_1,B_2 \leftrightarrow E$ & $A_1,A_2,B_1,B_2 \leftrightarrow E$ \\
$X,X^{\prime}$ & $A \leftrightarrow B_3; B_1 \leftrightarrow B_2$ & $A \leftrightarrow B_2; B_1 \leftrightarrow B_3$   \\
$Y,\Delta^{\prime}$ &  $A \leftrightarrow A; B \leftrightarrow B$ & $A \leftrightarrow B$   \\
$\Delta,Y^{\prime}$ &  $A \leftrightarrow B$ & $A \leftrightarrow A; B \leftrightarrow B$   \\
$\Sigma$ & $A,B \leftrightarrow A,B$ &  $A,B \leftrightarrow A,B$  \\
\bottomrule
\end{tabular*}
\label{tab:selection_rule}
\end{table*}

\subsection{Optical transition selection rules}
\subsubsection{Optical absorption spectrum and optical transitioin selectioin rules.}
The optical conductivity tensor $\sigma_{\alpha\beta}(\omega)$ is calculated using the Kubo formula:
\begin{equation}
\begin{split}
\sigma_{\alpha\beta}(\omega) = &\frac{\mathrm{i} e^2 \hbar}{N_k \Omega_c} \sum_{\mathbf{k}} \sum_{n,m} \frac{f_{m\mathbf{k}} - f_{n\mathbf{k}}}{\epsilon_{m\mathbf{k}} - \epsilon_{n\mathbf{k}}} \\
& \times \frac{\langle \psi_{n\mathbf{k}} | v_\alpha | \psi_{m\mathbf{k}} \rangle \langle \psi_{m\mathbf{k}} | v_\beta | \psi_{n\mathbf{k}} \rangle}{\epsilon_{m\mathbf{k}} - \epsilon_{n\mathbf{k}} - (\hbar \omega + \mathrm{i} \eta)},
\end{split}
\end{equation}
where $N_k$ is the number of $\mathbf{k}$-points sampling the BZ, $\Omega_c$ is the unit cell area, $\psi_{n\mathbf{k}}$ and $\epsilon_{n\mathbf{k}}$ denote the wave function and energy for the state with band index $n$ and wave vector $\mathbf{k}$, $f_{n\mathbf{k}}$ is the Fermi-Dirac distribution function, and $v_{\alpha}$ represents the $\alpha$-component of the velocity operator $\bm{v} = \frac{1}{\hbar} \nabla_{\mathbf{k}} H(\mathbf{k})$. The parameter $\eta \rightarrow 0^{+}$, and we adopt $\eta = 5$ meV in this paper. All optical conductivity calculations are carried out using the TBPLaS package~\cite{tbplas}.

In Fig.~\ref{fig:ac}, we show the optical absorption, namely the real part of the optical conductivity $\text{Re}[\sigma_{xx}]$, for the monolayer, AA-stacked, AB-stacked, and $36.87^\circ$ twisted bilayer in the photon energy range below $2.0$~eV. Only the $xx$ component is plotted, as the system symmetries ensure $\sigma_{xx} = \sigma_{yy}$. To interpret the optical response, we derive the OTSR using a standard group-theoretic approach. These rules determine whether the matrix element
\begin{equation}
\langle\varphi_f | v_{\alpha} | \varphi_i \rangle \neq 0
\end{equation}
is symmetry-allowed, where $\varphi_i$ and $\varphi_f$ denote the initial and final states during the optical transition, respectively. Only vertical transitions are considered, since the photon momentum is negligible. Selection rules can therefore be established independently at each $\bm{k}$-point in the BZ. From the relation $\hat{\bm{v}} = \frac{i}{\hbar}[\hat{H}, \hat{\bm{r}}]$, we derive the matrix element between initial and final states:
\begin{equation}
\begin{split}
\langle \varphi_f | \hat{\bm{v}} | \varphi_i \rangle &= \frac{i}{\hbar} \langle \varphi_f | [\hat{H}, \hat{\bm{r}}] | \varphi_i \rangle \\
&= \frac{i}{\hbar} (E_f - E_i) \langle \varphi_f | \hat{\bm{r}} | \varphi_i \rangle.
\end{split}
\end{equation}
Since $E_f \neq E_i$ for interband transitions, establishing the OTSR for $\alpha$-polarized light is equivalent to determining whether $\langle \varphi_f | r_{\alpha} | \varphi_i \rangle \neq 0$, which aligns with the standard dipole formulation in textbooks.

Let $ir_i$ and $ir_f$ denote the irreps of the initial state $\varphi_i$ and final state $\varphi_f$, respectively. For a given wavevector $\bm{k}$ and its little group, we consider the set $\{R r_{\alpha} \,|\, R \in \text{little group}\}$ formed by applying all little group operations to the position component $r_{\alpha}$. All the position components in this set as the basis functions construct a representation $\Gamma_v$ of the little group, which decomposes into the direct sum of irreps:
\begin{equation}
\Gamma_v \simeq \bigoplus_{\mu} a_{\mu} ir_{\mu},
\end{equation}
where $a_{\mu}$ is the multiplicity of irrep $ir_{\mu}$. The transition $ir_i \rightarrow ir_f$ is symmetry-allowed if the direct product representation $\Gamma_v \otimes ir_i$ contains $ir_f$.

Following this standard procedure, we derive the OTSRs for the monolayer, AA-stacked, and twisted bilayer at all high-symmetry points and along high-symmetry lines. These results are further validated by direct numerical evaluation of the matrix elements $\langle \varphi_f | v_{\alpha} | \varphi_i \rangle$. For the AB-stacked bilayer, which has a non-SSG due to the glide plane, the above procedure becomes more involved. Instead, we can determine its selection rules by explicitly computing the matrix elements numerically. The resulting OTSRs for all four structures are summarized in Table~\ref{tab:selection_rule}.

We now consider vertical transitions at a fixed wavevector. To facilitate the analysis, we introduce two sets of wavevectors: $\mathcal{LK}_1 = \{\Gamma, M, \Sigma\}$ and $\mathcal{LK}_2 = \{X, X^{\prime}, Y, Y^{\prime}, \Delta, \Delta^{\prime}\}$ for the monolayer, AA-stacked bilayer, and twisted bilayer; and $\mathcal{LK}_1 = \{\Gamma, M, \Sigma, Y, Y^{\prime}\}$ and $\mathcal{LK}_2 = \{X, X^{\prime}, \Delta, \Delta^{\prime}\}$ for the AB-stacked bilayer. For $\bm{k} \in \mathcal{LK}_1$, the OTSRs are identical for $x$- and $y$-polarized light in all systems under consideration. In contrast, for $\bm{k} \in \mathcal{LK}_2$, the system exhibits distinct OTSRs for $x$- and $y$-polarized light: if a transition is allowed for one polarization, it is forbidden for the other. Hence, for wavevectors in $\mathcal{LK}_2$, the OTSR becomes polarization-dependent.

Comparing the OTSRs for $\bm{k} \in \mathcal{LK}_2$ and its symmetric counterpart $\bm{k}^{\prime} = \sigma_{(xy)z}\bm{k}$ reveals that if a vertical transition at $\bm{k}$ is allowed by $x$- or $y$-polarized light, the equivalent vertical transition at $\bm{k}^{\prime}$ is allowed by another polarization. For $\bm{k} = Y$, $Y^{\prime}$, $\Delta$, and $\Delta^{\prime}$, this behavior can be derived directly from Table~\ref{tab:selection_rule}. However, for the monolayer, AA-stacked bilayer, and twisted bilayer, Table~\ref{tab:selection_rule} appears to suggest that the equivalent transitions at $X$ and $X^{\prime}$ are allowed by the same polarization. We now clarify that $X$ and $X^{\prime}$ are not exceptions to the rule; indeed, the two equivalent transitions are also allowed by different polarizations. For these systems, any allowed transition at $X$ or $X^{\prime}$ according to Table~\ref{tab:selection_rule} involves an energy level whose irrep changes under the $\sigma_{(xy)z}$ operation. Consider a transition at $\bm{k} = X$ (or $X^{\prime}$) expressed as $ir \rightarrow ir^{\prime}$, where an electron is excited from an initial state with irrep $ir$ to a final state with irrep $ir^{\prime}$. Under the symmetry operation, we assume that the final-state irrep transforms to $ir^{\prime\prime}$ at the equivalent wavevector $\bm{k}^{\prime} = \sigma_{(xy)z}\bm{k}$, so that the transition $ir \rightarrow ir^{\prime}$ at $\bm{k}$ becomes to $ir \rightarrow ir^{\prime\prime}$ at $\bm{k}^{\prime}$. Table~\ref{tab:selection_rule} then shows that if the transition $ir \rightarrow ir^{\prime}$ at $\bm{k}$ is allowed by $x$- or $y$-polarized light, the corresponding transition $ir \rightarrow ir^{\prime\prime}$ at $\bm{k}^{\prime}$ is allowed by the opposite polarization.

\subsubsection{Understanding optical absorption spectra using optical transition selection rules.}
The optical absorption spectrum for specifically polarized light is determined by both the OTSRs and the Pauli exclusion principle. In this subsection, we apply these two fundamental rules to understand the key features of the optical absorption spectrum for photon energies below 2 eV for the four systems, specifically focusing on the emergence of absorption steps and peaks. In Fig.~\ref{fig:ac}, these features are indicated for the four systems: starting points of absorption steps are marked by vertical gray dashed lines, and absorption peaks by vertical green dashed lines. These features are associated with the vertical transitions at high-symmetry points and along symmetry lines in the BZ. Furthermore, we restrict our analysis to $Re[\sigma_{xx}]$ for $x$-polarized light, as the method is directly applicable to $Re[\sigma_{yy}]$ for $y$-polarized light and the latter does not introduce any new difference in physics.

Generally, an absorption step appears when a new interband transition channel becomes active. The onset energy of this step corresponds to the minimum energy required for this interband transition. The first absorption step in energy, therefore, marks the beginning of non-zero optical absorption. Absorption peaks, on the other hand, originate from transitions between DOS peaks. In the following, we provide a detailed explanation of the absorption steps and peaks for the four systems.

\textit{Monolayer.---} 
For the monolayer, the optical absorption spectrum features a distinct peak at 1.37\,eV and an absorption step starting at 0.38\,eV. According to the Pauli exclusion principle, only the transitions from the highest VB to the two lowest CBs, labeled by $V_1 \rightarrow C_1$ and $V_1 \rightarrow C_2$, contribute to the low-energy optical absorption. Considering the OTSR, the $V_1 \rightarrow C_1$ transition is allowed only for wavevectors along the path $M-\Gamma-X^{\prime}-M$ in the BZ for $x$-polarized light. For brevity, we denote this interband transition allowed by $x$-polarized light as $V_1 \rightarrow C_1: M-\Gamma-X^{\prime}-M$, which explicitly specifies the allowed wavevector region. Similarly, the $V_1 \rightarrow C_2$ transition is permitted along the path $\Gamma-X-M-\Gamma$, denoted as $V_1 \rightarrow C_2: \Gamma-X-M-\Gamma$. Both interband transitions share the same minimum transition energy of 0.38\,eV, which corresponds to the fundamental band gap and results in the absorption step starting at that energy. The optical absorption peak at 1.37\,eV arises from the $V_1 \rightarrow C_1: M-\Gamma-X^{\prime}-M$ interband transition at the $X^{\prime}$ point labeled by blue arrow in Fig. \ref{fig:band_irrep}(a), because this vertical transition correspond to the transition between DOS peaks.

\textit{AA-stacked bilayer.---} 
For the AA-stacked bilayer, the low-energy optical absorption spectrum in Fig.~\ref{fig:ac}(b) reveals three obvious absorption steps and two absorption peaks. Notably, the optical absorption from the first step nearly vanishes before the onset of the second step. Guided by the Pauli exclusion principle and the OTSRs, the interband transitions contributing to the low-energy optical absorption include:
\begin{enumerate}
    \item $V_{1+} \rightarrow C_{1+}: M-\Gamma-X^{\prime}-M$, excluding the region around $M$ where both bands lie below the Fermi energy. The minimum transition energy is 0.61\,eV [see Fig.~\ref{fig:band_irrep}(e)]. A peak appears at 1.47\,eV, arising from the vertical transitions at $X^{\prime}$, corresponding to the transition between DOS peaks as shown in Fig.~\ref{fig:band_irrep}(b).
    \item $V_{1+} \rightarrow C_{2+}: \Gamma-X-M-\Gamma$, excluding the region around $M$ where both bands are below the Fermi energy. The minimum transition energy is 0.48\,eV [see Fig.~\ref{fig:band_irrep}(e)]. This interband transition does not produce an absorption peak below 2\,eV.
    \item $V_{1-} \rightarrow C_{1-}: M-\Gamma-X^{\prime}-M$, excluding the region around $M$ where both bands are above the Fermi energy. The minimum transition energy is 0.75\,eV [see Fig.~\ref{fig:band_irrep}(f)]. This transition at $X^{\prime}$ is responsible for the absorption peak at 1.27\,eV, corresponding to the transition between DOS peaks as shown in Fig.~\ref{fig:band_irrep}(b).
    \item $V_{1-} \rightarrow C_{2-}: \Gamma-X-M-\Gamma$, excluding the region around $M$ where both bands are above the Fermi energy. The minimum transition energy is 1.04\,eV [see Fig.~\ref{fig:band_irrep}(f)]. This interband transition does not produce an absorption peak below 2\,eV.
    \item $C_{1+} \rightarrow C_{2+}: X^{\prime}-M-\Gamma$, including only the region around $M$ where $C_{1+}$ is below and $C_{2+}$ is above the Fermi energy, as depicted in Fig.~\ref{fig:band_irrep}(g). The transition energies for this channel range from the minimum of 0.025\,eV to the maximum of 0.35\,eV.
\end{enumerate}
From this analysis, we can attribute the absorption peaks at 1.47 and 1.27\,eV to interband transitions 1 and 3, respectively. The absorption steps starting at approximately 0.02, 0.48, and 0.75\,eV originate from interband transitions 5, 2, and 3, respectively. The minimum transition energy of interband transition 1 (0.61\,eV) does not manifest as a discernible step, while the onset of interband transition 4 (1.04\,eV) appears as a weak step in Fig.~\ref{fig:ac}(b). An interband transition contributes to absorption across a continuous energy range, from its minimum to its maximum transition energy. Consequently, the entire absorption feature in the range of $0.02 \sim 0.35$\,eV is attributed solely to interband transition 5. Furthermore, the absorption from this channel diminishes to nearly zero just before the onset of the next absorption step starting at 0.48\,eV.

\textit{AB-stacked bilayer.---} 
Due to weak interlayer coupling, the AB-stacked bilayer exhibits an optical absorption spectrum similar to that of the monolayer. The key differences are two absorption steps, starting at 0.20 and 0.55\,eV, and a single absorption peak at 1.34\,eV. Applying the selection rules and Pauli exclusion principles, the contributing interband transitions are:
\begin{enumerate}
    \item $V_1 \rightarrow C_1: X-M-\Gamma-X^{\prime}-M$. The minimum transition energy is 0.20\,eV at the $M$ point. The absorption peak at 1.34\,eV originates from transitions on the $\Delta^{\prime}$ line near $X^{\prime}$, corresponding to the transition between DOS peaks as shown in Fig.~\ref{fig:band_irrep}(c).
    \item $V_2 \rightarrow C_2: X-M-\Gamma-X^{\prime}-M$. The minimum transition energy is also 0.20\,eV at the $M$ point.
    \item $V_1 \rightarrow C_3: X^{\prime}-M$ and $\Gamma-X-M-\Gamma$. The minimum transition energy is 0.55\,eV at the $M$ point.
    \item $V_2 \rightarrow C_4: X^{\prime}-M$ and $\Gamma-X-M-\Gamma$. The minimum transition energy is 0.55\,eV at the $M$ point.
\end{enumerate}
Therefore, the absorption peak at 1.34\,eV is attributed to interband transition 1. The absorption step starting at 0.20\,eV arises from the combined onsets of transitions 1 and 2, while the step starting at 0.55\,eV originates from the combined onsets of transitions 3 and 4.

\begin{figure*}[tbp]
\centering
\includegraphics[width=0.8\textwidth]{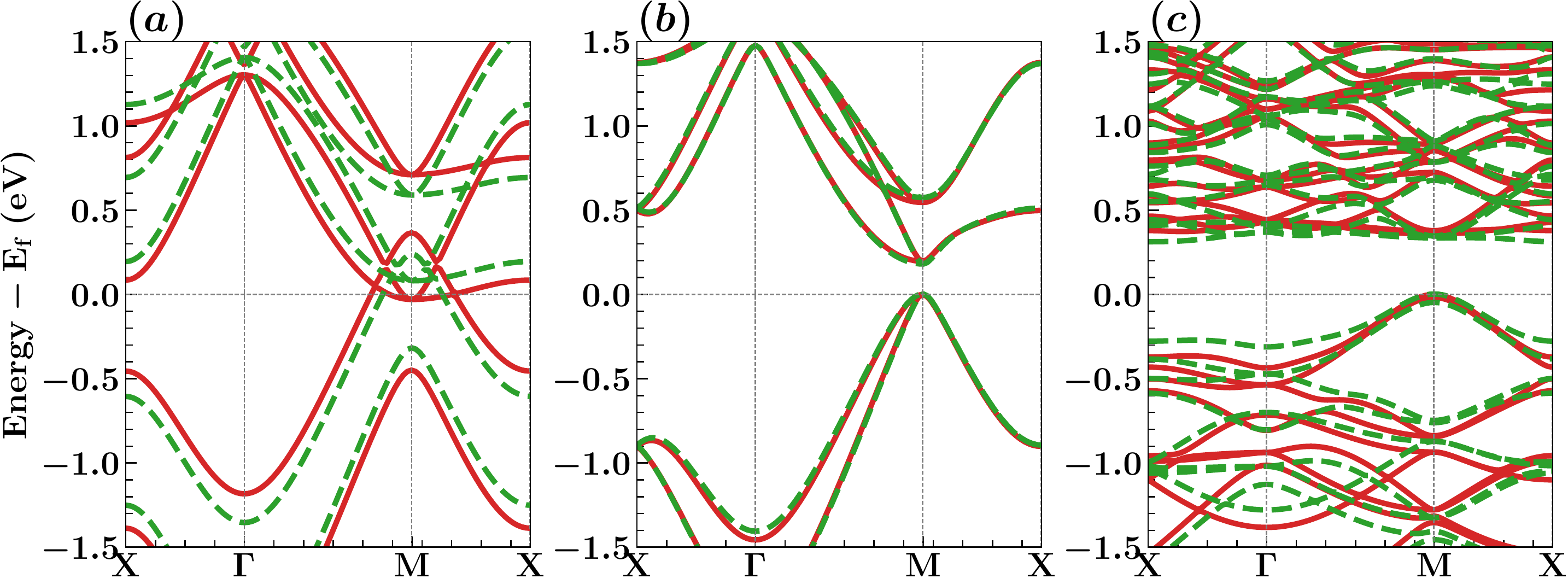}
\caption{Band structure comparison between rigid and relaxed structures for (a) AA-stacked, (b) AB-stacked, and (c) 36.87$^{\circ}$ twisted bilayer. Solid red and dashed green lines are the band structures of rigid and relaxed structures, respectively. The Fermi energy of the rigid structure is set to 0 eV.}
\label{fig:band_comp_relax}
\end{figure*}

\textit{Twisted bilayer.---} Table \ref{tab:selection_rule} shows that optical transitions in the $36.87^\circ$ twisted bilayer remain polarization-dependent at high-symmetry points $X$, $X^{\prime}$ and along symmetry lines $\Delta$, $\Delta^{\prime}$, $Y$, and $Y^{\prime}$. However, the twisted configuration produces a high density of closely spaced bands, making it less meaningful to focus on the polarization dependence of individual transitions, because one can typically find other transitions with similar energy excited by the oppositely polarized light at the same position in the BZ. For example, as shown in Fig.~\ref{fig:band_irrep}(d), the transition from the highest VB to the lowest CB at $X$ ($B_1\rightarrow B_2$) can be excited by only $x$-polarized light, and the equivalent transition at $X^{\prime}$ ($B_1\rightarrow B_3$) can be excited by only $y$-polarized light. Simultaneously, the transition with similar energy but excited by the light with the opposite polarization occur at the same positions: at $X$, the transition from the second highest VB to the lowest CB ($A\rightarrow B_2$) can be excited by only $y$-polarized light, while at $X^{\prime}$, the equivalent transition ($A\rightarrow B_3$) can be excited by $x$-polarized light.

Fortunately, despite this complexity, we can still correlate optical absorption peaks with transitions between DOS peaks by comparing Figs.~\ref{fig:band_irrep}(d) and \ref{fig:ac}(d). Denoting the three main DOS peaks below the Fermi energy as $p_{1-}$, $p_{2-}$, $p_{3-}$ and the two above as $p_{1+}$, $p_{2+}$, we attribute the optical absorption peaks at 0.78 eV, 0.91 eV, 1.01 eV, 1.17 eV, and 1.33 eV to transitions between the following DOS peaks: $p_{1-}\rightarrow p_{1+},\  p_{1-}\rightarrow p_{2+},\ p_{2-}\rightarrow p_{1+},\ p_{2-}\rightarrow p_{2+}$, and $p_{3-}\rightarrow p_{2+}$, respectively.

\subsection{Relaxation effect}
When two layers of a 2D material are stacked, the resulting bilayer structure typically undergoes atomic relaxation to minimize its total energy. Here, we investigate how such structural relaxation influences the electronic and optical properties of bilayer ZnPc-MOF. To account for this effect, we fully relax the AA-stacked, AB-stacked, and 36.87$^{\circ}$ twisted bilayer structures using the VASP code\cite{vasp}. All calculations are performed within the PBE functional\cite{pbe} including vdW correction\cite{vdw0}. A plane-wave kinetic energy cutoff of 520 eV is applied, and vacuum layers of approximately 20 {\AA} are introduced to avoid spurious interactions between periodic images. For the AA- and AB-stacked bilayers, we employ a $5\times5\times1$ Gamma-centered $k$-point mesh, while a $3\times3\times1$ mesh is used for the 36.87$^{\circ}$ twisted bilayer. The Gaussian smearing width is set to 0.05 eV. The atomic positions are optimized until the residual forces on each atom fall below 0.01 eV/{\AA}.

After relaxation, the average interlayer distances, measured using the carbon and nitrogen atoms as the reference that dominate the electronic states near the Fermi level, are 3.52, 3.31, and 3.21 {\AA} for the AA-stacked, AB-stacked, and 36.87$^{\circ}$ twisted bilayers, respectively. The constituent layers exhibit slight deviations from perfect planarity. The vertical fluctuation, defined as the difference between the highest and lowest $z$-coordinates within each monolayer, amounts to 0.02, 0.09, and 0.19 {\AA} for the three systems. Thus, the AA-stacked bilayer remains nearly flat, whereas the twisted bilayer shows the most pronounced out-of-plane distortion.

The band structures of the three relaxed bilayers are presented in Fig.~\ref{fig:band_comp_relax}. Compared to the rigid models with fixed interlayer distance of 3.38 {\AA} yielding band gaps of -0.4, 0.2, and 0.36 eV, respectively, the relaxed AA-stacked bilayer exhibits an increased interlayer distance and a correspondingly larger band gap of -0.16 eV. In contrast, the AB-stacked and 36.87$^{\circ}$ twisted bilayers show reduced interlayer distances and narrower band gaps of 0.18 and 0.33 eV. These trends reflect a weakening of interlayer coupling in the AA-stacked system and a strengthening in the other two systems. In addition, in the 36.87$^{\circ}$ twisted bilayer, relaxation reconstructs strongest the band structure among three bilayers, such as the flattening of the low-energy VBs.

Figure \ref{fig:ac} displays the optical absorption spectra of the three relaxed bilayers. Owing to the relatively small change in interlayer distance upon relaxation, the absorption spectra of the AB-stacked bilayer is the least affected. In the AA-stacked case, the reduced interlayer coupling modifies the minimum transition energies of interband transitions 1$\sim$4 and alters both the minimum and maximum transition energies of interband transition 5. Although these changes lead to some fine-tuning of the spectral features, all key characteristics, including the absorption steps and peaks, remain intact. Among the three systems, the 36.87$^{\circ}$ twisted bilayer experiences the most significant modification to its absorption spectra, driven by the relaxation-induced band restructuring. Nevertheless, the onset of absorption and the multi-peak structure are preserved.

Importantly, the space group symmetries of all three systems are retained after relaxation, ensuring that our symmetry- and group-theory-based analyses remain valid.

\subsection{Discussion}
In this paper, we employ the $p_z$-orbital-based tight-binding model to investigate the electronic structures and optical properties of ZnPc-MOF systems. For the monolayer, the unit cell contains 28 $p_z$ orbitals, leading to a 28-band model. Notably, only three isolated bands appear near the Fermi energy within the energy range of $-1.5$ to $1.5$ eV, consisting of one VB and two CBs. This feature suggests the possibility of constructing a reduced tight-binding model to capture the low-energy electronic behavior. A natural basis choice might involve the molecular orbitals of phthalocyanine; however, the HOMO and LUMO are not suitable for this purpose, as they lie far from the Fermi energy, whereas our focus is on the electronic states in the vicinity of the Fermi level. Instead, an optimal approach is to construct the maximally localized Wannier functions (MLWFs)\cite{MLWF_rmp} obtained from the three bands around the Fermi energy, thereby enabling a three-band model for the ZnPc-MOF monolayer. Nevertheless, extending such a model to bilayers with various stacking configurations requires a general interlayer hopping function between MLWFs, which is a significant challenge. Moreover, if relaxation effects induce the fluctuations, this model may become inadequate and necessitate corrections. In contrast, the $p_z$-orbital-based tight-binding model remains applicable to multilayer structures with any interlayer stacking even in the presence of relaxation. Therefore, the model adopted in this work offers an optimal balance between computational accuracy and efficiency, making it well-suited for capturing the electronic structures and optical properties of ZnPc-MOF systems.

\begin{figure}[!htbp]
\centering
\includegraphics[width=6 cm]{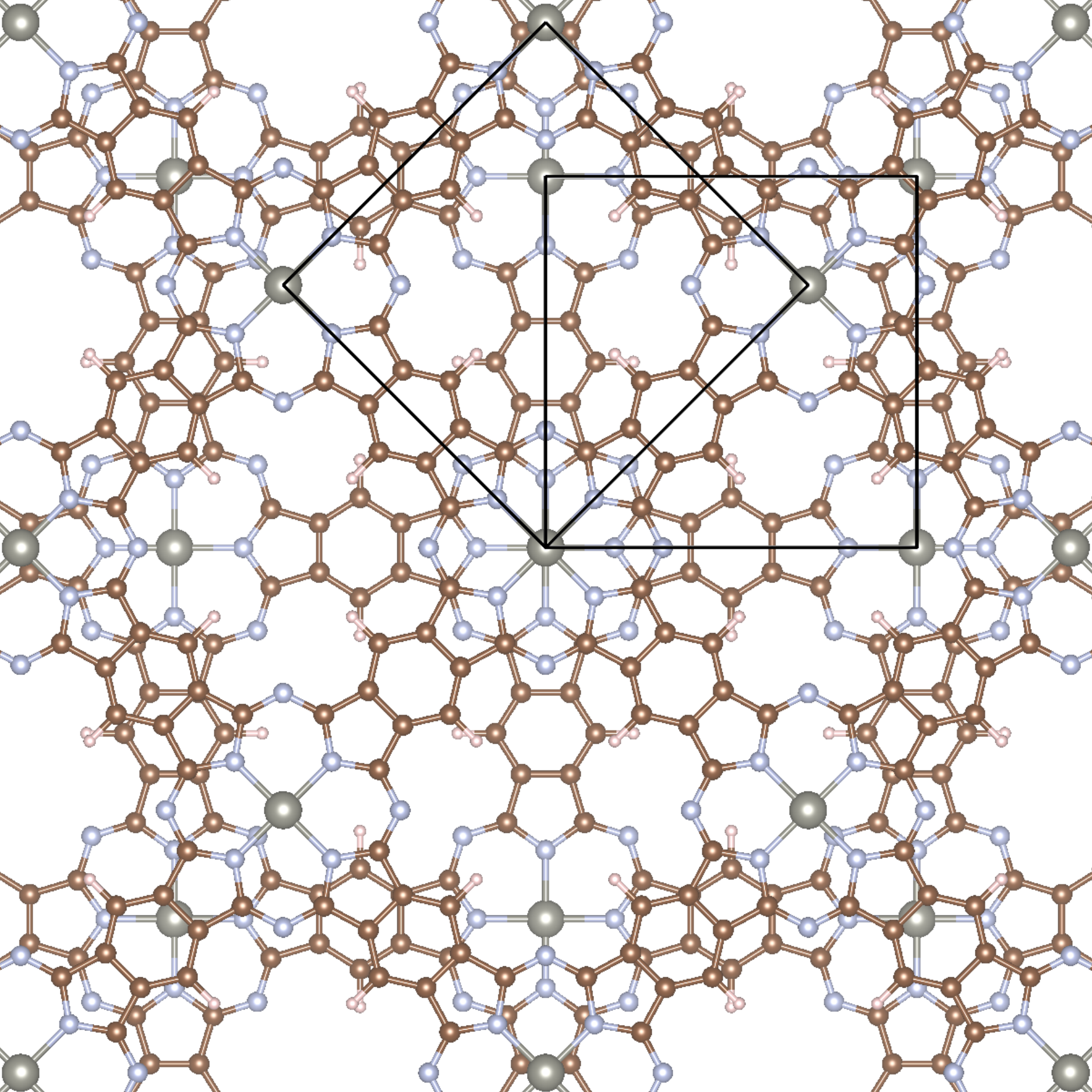}
\caption{The structure of 45$^{\circ}$ twisted bilayer. The unit cells of two layers are shown.}
\label{fig:st_qc}
\end{figure}

\begin{figure}[!htbp]
\centering
\includegraphics[width=8.5 cm]{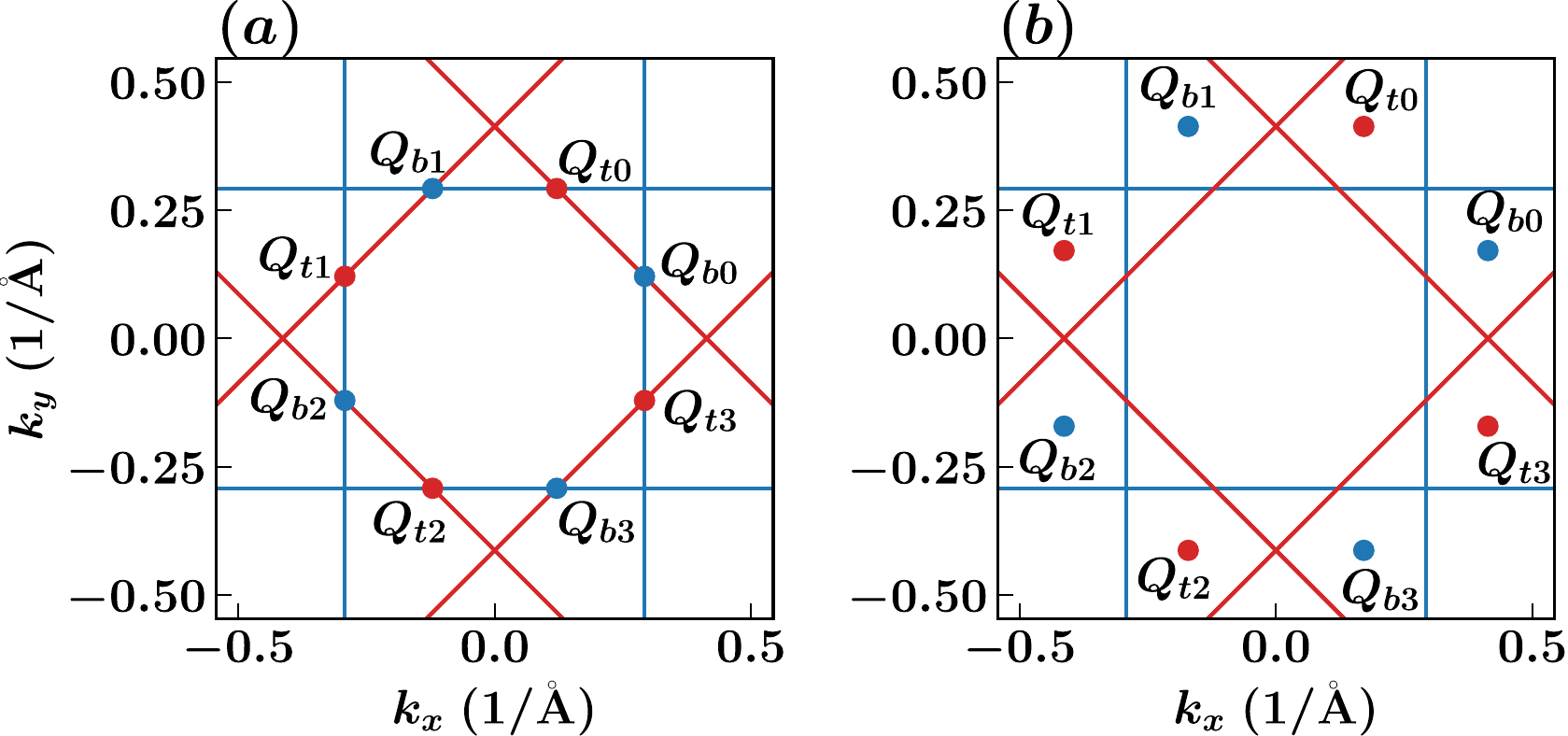}
\caption{The positions of $\bm{Q}_{bi}$'s and $\bm{Q}_{ti}$'s for (a) strongest and (b) second strongest resonant couplings.}
\label{fig:bases}
\end{figure}

\begin{figure*}[!htbp]
\centering
\includegraphics[width=16 cm]{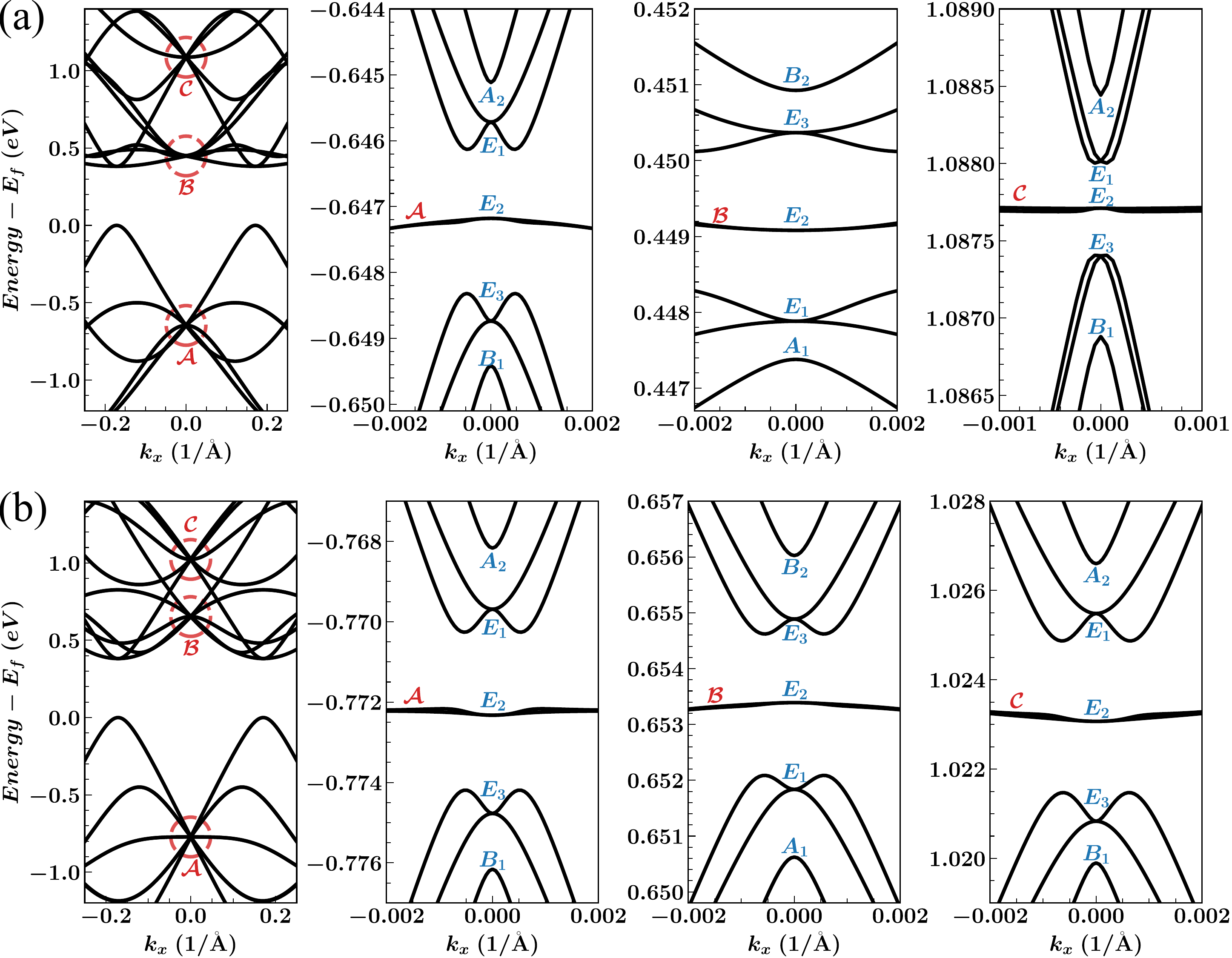}
\caption{The quasi-band structures for (a) strongest and (b) second strongest resonant couplings. The three parts around $\Gamma$ for each resonant coupling (labeled by $\mathcal{A}$, $\mathcal{B}$, and $\mathcal{C}$) are zoomed in the last three columns. The irreps of the quasicyrstalline electronic states are marked.}
\label{fig:bands_qc}
\end{figure*}

\begin{figure*}[!htbp]
\centering
\includegraphics[width=16 cm]{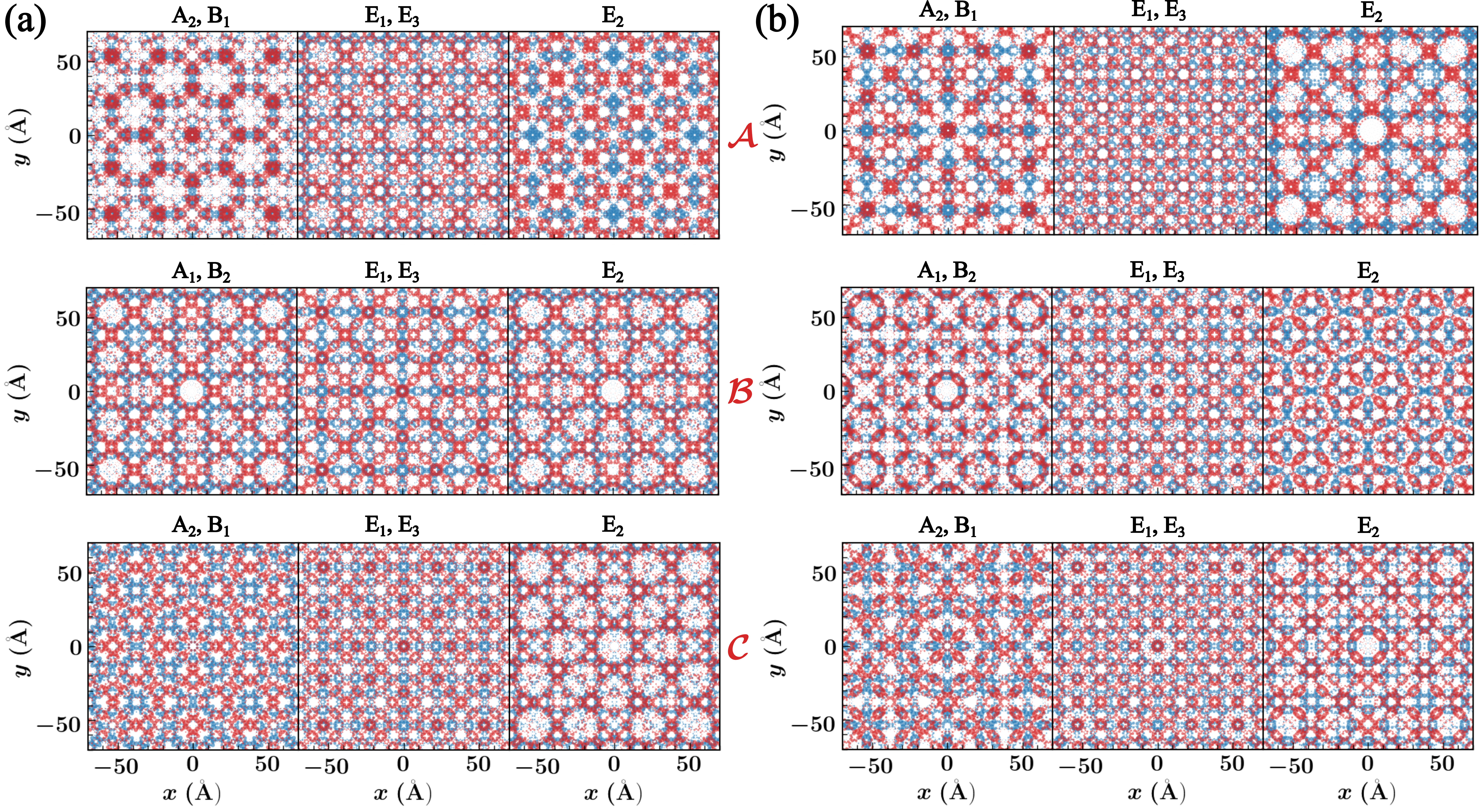}
\caption{The quasicrystalline electronic states for (a) strongest and (b) second strongest resonant couplings. The three rows show the quasicrystalline states marked by $\mathcal{A}$, $\mathcal{B}$, and $\mathcal{C}$ as given in Fig. \ref{fig:bands_qc}. The dot size stands for the occupation of the state on this site, and the occupations on the bottom and top layers are plotted in blue and red, respectively.}
\label{fig:qc_states}
\end{figure*}

\section{Quasicrystalline electronic states of 45$^{\circ}$ twisted bilayer}
When the twist angle is $45^{\circ}$, the twisted bilayer forms an octagonal vdW quasicrystal with broken translational symmetry. In this case, the system exhibits point group symmetry $\mathbb{D}_{4d}$, and conventional band theory is no longer applicable. Instead, we employ the k-space tight-binding model as a theoretical framework to investigate its electronic structure, following the approach previously established for graphene quasicrystal~\cite{kTB0,kTB1}.
\subsection{k-space tight-binding model and resonant coupling Hamiltonians}
Within this framework, we use the Bloch functions of each layer as basis functions:
\begin{equation}
\begin{split}
\phi_{\bm{k}_l,X}(\bm{r}) &= \langle \bm{r} |\bm{k}_l,X;l \rangle \\
&= \frac{1}{\sqrt{N_l}} \sum_{\bm{R}_l} e^{i\bm{k}_l\cdot(\bm{R}_l+\bm{\tau}_{X}^l)} p_z(\bm{r}-\bm{R}_l-\bm{\tau}_{X}^l),
\end{split}
\end{equation}
where $N_l$ is the number of unit cells in layer $l$, $\bm{R}_l$ denotes the lattice vectors of layer $l$, and $\bm{\tau}_X^l$ represents the position of sublattice $X$ within the unit cell of layer $l$. The intralayer Hamiltonian matrix elements are given by:
\begin{equation}
\begin{split}
&\langle\bm{k}_l,X;l|H|\bm{k}_l^{\prime},X^{\prime};l\rangle = h_{XX^{\prime}}^l(\bm{k}_l)\delta_{\bm{k}_l,\bm{k}_l^{\prime}},\\
&h_{XX^{\prime}}^l(\bm{k}_l) = \sum_{\bm{R}_l}t(\bm{R}_l+\bm{\tau}_{X^{\prime}}^l-\bm{\tau}_X^l)e^{i\bm{k}_l\cdot(\bm{R}_l+\bm{\tau}_{X^{\prime}}^l-\bm{\tau}_X^l)}.
\end{split}
\end{equation}
The interlayer Hamiltonian matrix elements take the form~\cite{interlayer_rule_0,interlayer_rule_1}:
\begin{equation}
\begin{split}
\langle\bm{k}_b,X;b|U&|\bm{k}_t,X^{\prime};t\rangle = \sum_{\bm{G}_b,\bm{G}_t} T(\bm{k}_b+\bm{G}_b) \\
\times & e^{i\bm{G}_b\cdot\bm{\tau}_{X}^b - i\bm{G}_t\cdot\bm{\tau}_{X^{\prime}}^t} \delta_{\bm{k}_b+\bm{G}_b,\bm{k}_t+\bm{G}_t},
\end{split}
\label{eq:intercoup_condition}
\end{equation}
where $T(\bm{k})$ is the in-plane Fourier transform of the interlayer hopping function $t(\bm{r})$:
\begin{equation}
T(\bm{q}) = \frac{1}{S}\int t(\bm{r}_{xy}+h\hat{\bm{e}}_z) e^{-i\bm{q}\cdot\bm{r}_{xy}}d\bm{r}_{xy}.
\end{equation}
Here, $S$ is the area of the monolayer unit cell and $h$ is the interlayer distance. For our tight-binding model, $T(\bm{q})$ depends only on the magnitude of $\bm{q}$, i.e., $T(\bm{q})=T(|\bm{q}|)$. Equation~\ref{eq:intercoup_condition} establishes the interlayer coupling condition, that is Bloch functions from different layers can couple with each other when $\bm{k}_b+\bm{G}_b=\bm{k}_t+\bm{G}_t$ is satisfied.

A defining characteristic of vdW quasicrystals, distinct from other bilayer structures, is the existence of an infinite number of quasicrystalline electronic states. These states emerge from an infinite series of resonant coupling Hamiltonians~\cite{Moon_resonant}, each characterized by a unique choice of basis functions (in detail the choice of wavevectors of two layers $\bm{k}_b$ and $\bm{k}_t$), corresponding to a specific resonant coupling strength. In this work, we focus on the two strongest resonant couplings. For each case, the wavevectors of the two layers are restricted to $\bm{k}_b=\bm{k}_0+\bm{Q}_{bi}$ ($i=0$--$3$) and $\bm{k}_t=\bm{k}_0+\bm{Q}_{ti}$ ($i=0$--$3$). The positions of the vectors $\bm{Q}_{bi}$ and $\bm{Q}_{ti}$ for the two strongest couplings~\cite{2strongest0,2strongest1} are shown in Fig.~\ref{fig:bases}(a) and (b), respectively. For any pair $\bm{k}_b=\bm{k}_0+\bm{Q}_{bi}$ and $\bm{k}_t=\bm{k}_0+\bm{Q}_{ti}$, there exists exactly one pair of reciprocal lattice vectors $\bm{G}_b$ and $\bm{G}_t$ that satisfies the coupling condition $\bm{k}_b+\bm{G}_b=\bm{k}_t+\bm{G}_t=\bm{q}$; this uniqueness is guaranteed by the incommensurate twist angle. The resonant coupling strength is defined as the maximum value of $T(|\bm{q}|)$ among all such wavevector pairs at $\bm{k}_0=0$. For the two strongest couplings considered here, the strengths are 35.4~meV and 33~meV, respectively.

To compute the electronic structure of the ZnPc-MOF quasicrystal, we first construct the $\bm{k}_0$-dependent resonant coupling Hamiltonian $H(\bm{k}_0)$. Diagonalization of $H(\bm{k}_0)$ yields the energy dispersion $E(\bm{k}_0)$, referred to as the quasi-band structure following the terminology introduced by P.~Moon for graphene quasicrystals\cite{Moon_resonant}. The exact quasicrystalline electronic states are obtained at $\bm{k}_0=0$; as $\bm{k}_0$ moves away from the $\Gamma$ point, these states gradually deviate from the ideal quasicrystalline character. The resulting quasi-band structures for the two strongest resonant couplings are presented in Fig.~\ref{fig:bands_qc}. 

\subsection{Quasicrystalline electronic states and symmetry}
At $\bm{k}_0=0$, the Hamiltonian $H(\bm{k}_0)$ commutes with all operations of the $\mathbb{D}_{4d}$ point group, allowing the quasicrystalline electronic states to be classified according to the irreps of $\mathbb{D}_{4d}$ as given in Appendix \ref{app:tables}. Using the Bloch functions of each layer as basis functions, the representation matrix elements for a symmetry operation $R$ are given by:
\begin{equation}
\begin{split}
D_{l,\bm{k}_l,X_1}^{l^{\prime},\bm{k}_{l^{\prime}},X_2}(R) &\equiv\langle\bm{k}_l,X_1;l |P_R|\bm{k}_{l^{\prime}},X_2;l^{\prime}\rangle \\
&= \sum_{\bm{G}_{l_R^{\prime}}} \zeta e^{i\bm{G}_{l_R^{\prime}}\cdot\bm{\tau}_{X_2^{\prime}}} \delta_{l,l_R^{\prime}} \delta_{\bm{k}_l+\bm{G}_{l_R^{\prime}},R\bm{k}_{l^{\prime}}} \delta_{X_1,X_2^{\prime}},
\end{split}
\end{equation}
where $R\bm{\tau}_{X_2}=\bm{R}_{l_R^{\prime}}+\bm{\tau}_{X_2^{\prime}}$, and $l_R^{\prime}=l^{\prime}$ if $R$ preserves the layer index and $l_R^{\prime}=\overline{l^{\prime}}$ (the complementary layer) if $R$ exchanges layers. We then construct the projection operator matrix following Eq.~\eqref{eq:proj_op} to identify the irreps of the eigenstates of $H(\bm{k}_0)$. The resulting irrep assignments for the quasicrystalline electronic states are labeled in Fig.~\ref{fig:bands_qc}. States belonging to 1D irreps ($A_1$, $A_2$, $B_1$, $B_2$) are simultaneously eigenstates of all symmetry operations, while those in 2D irreps ($E_1$, $E_2$, $E_3$) are not. Nevertheless, we can identify states that are simultaneous eigenstates of both the Hamiltonian and one specific symmetry operation such as $S_8$~\cite{interlayer_rule_Yu}.

Figure~\ref{fig:qc_states} displays the spatial distributions of quasicrystalline electronic states for both the strongest and second strongest resonant couplings. These states are also eigenstates of the $S_8$ operation, satisfying $S_8|\phi\rangle=\lambda|\phi\rangle$ with eigenvalues: $\lambda=1$ for $A_1$ and $A_2$; $\lambda=-1$ for $B_1$ and $B_2$; $\lambda=e^{\pm i\pi/4}$ for $E_1$; $\lambda=e^{\pm i\pi/2}$ for $E_2$; and $\lambda=e^{\pm i3\pi/4}$ for $E_3$. Notably, $A_1$ and $B_2$ states exhibit identical real-space patterns, as do $A_2$ and $B_1$ states, and similarly for $E_1$ and $E_3$ states.

\subsection{Discussion}
Searching for the vdW quasicrystal, whose quasicrystalline electronic states are stable and lie near the Fermi energy, is one of the main directions in vdW quasicrystal research. Such states contribute to the low-energy electronic structure; otherwise, the system would only exhibit quasicrystalline characteristics in its only atomic arrangement rather than in its electronic properties. In Table~\ref{tab:comp_QCs}, we compare the resonant coupling strength and the minimum distance from the quasicrystalline states to the band gap (or Fermi energy) between the ZnPc-MOF quasicrystal and graphene quasicrystal, the first experimentally realized and extensively studied vdW quasicrystal. Our calculations show that the quasicrystalline electronic states in the ZnPc-MOF quasicrystal lie closer to the band gap than those in graphene quasicrystals lie to the Fermi level at the Dirac point. This implies a greater involvement of quasicrystalline states in low-energy electronic phenomena. However, the resonant coupling strength in ZnPc-MOF quasicrystal is much smaller than that in graphene quasicrystal, so the quasicrystalline electronic states are not as stable as those in the graphene quasicrystal.

The behavior, that the quasicrystalline electronic states is closer to the band gap than graphene quasicrystal, is attributed to the smaller dispersion of the CBs in ZnPc-MOF. In the absence of interlayer coupling, the eight $Q$ points shown in Fig.~\ref{fig:bases} share exactly the same energy, which is closer to the band gap because of the small band dispersion. Upon introducing interlayer coupling, resonant coupling among these eight points generates new hybridized states whose energies are pushed closer to the band gap further.

Another important characteristic of quasicrystalline electronic states is their critical nature, meaning that they exhibit power-law localization instead of exponential localization. So these states extend over a large real-space region. However, we must point out that the $k$-space tight-binding model cannot capture their localization behavior, although it can capture the energy spectrum and even the energetic positions and symmetries of these quasicrystalline electronic states. The standard approach for obtaining the exact localization behavior is to diagonalize the Hamiltonian of a quantum-dot flake with a radius of tens of nanometers, but such diagonalization requires substantial computational resources and time that we cannot afford. Therefore, we do not explore the localization behavior of the quasicrystalline electronic states in the ZnPc-MOF quasicrystal in this paper.

In this work, we do not perform structural relaxation for the ZnPc-MOF quasicrystal, primarily due to the loss of periodicity. One possible approach to address this would be to use a quantum-dot flake containing tens of millions of atoms, similar to the method required to capture the localization behavior of quasicrystalline electronic states. However, relaxation involving such a large number of atoms is beyond the reach of DFT calculations. A more feasible alternative is to employ a classical force field, which can be implemented using the LAMMPS code\cite{lammps}, although the selection of an appropriate force field must be made with great care. Such an investigation is left for future work. Nevertheless, we expect that the $D_{4d}$ symmetry of the ZnPc-MOF quasicrystal is preserved, given that the symmetries of all other ZnPc-MOF bilayer structures and bilayer graphene systems including graphene quasicrystal are maintained after relaxation. Therefore, our symmetry-based analysis of the electronic structure of the ZnPc-MOF quasicrystal should remain valid.

\begin{table}[htb]
\caption{Comparison of quasicrystalline electronic states between ZnPc-MOF quasicrystal and graphene quasicrystal. $\mathcal{T}$ is the resonant coupling strength, and $\mathcal{D}$ denotes the minimum distance of the quasicrystalline electronic states to the band gap for ZnPc-MOF quasicrystal and to the Fermi energy for graphene quasicrystal. 1st and 2nd stand for the strongest and second strongest resonant coupling. The results for graphene quasicrystal are from Ref. \cite{Moon_resonant}.}
\centering
\begin{tabular*}{\linewidth}{@{\extracolsep{\fill}}lcccc@{}}
\toprule
 & \multicolumn{2}{c}{ZnPc-MOF} & \multicolumn{2}{c}{Graphene} \\
 & 1st & 2nd & 1st &2nd \\ 
\midrule
$\mathcal{T}$ (meV) & 35.4 & 33 & 157 & 48 \\
$\mathcal{D}$ (eV) & 0.07 & 0.27 & 1.58 & 3.29\\
\bottomrule
\end{tabular*}
\label{tab:comp_QCs}
\end{table}

\section{Conclusions}
Using the 2D ZnPc-MOF as a prototypical square-lattice system, we have systematically investigated the electronic structures and optical properties of its monolayer, AA- and AB-stacked bilayers, and twisted bilayers. By employing symmetry and group-theoretical analysis, we have classified the electronic bands according to the irreps of the little groups. A key finding is the two-fold degeneracy of the bands in the AB-stacked bilayer along the $Y$ and $Y^{\prime}$ lines, which originates from the existence of only one 2D irrep along these high-symmetry directions, a distinctive feature uncommon in hexagonal 2D materials. From the OTSRs, we have identified polarization-dependent optical transitions across the BZ. All symmetry-based analyses remain valid after structural relaxation and are applicable to other 2D square lattices sharing the same space group. Furthermore, we find that the quasicrystalline electronic states in the $45^{\circ}$ twisted bilayer lie closer to the Fermi level than those in graphene quasicrystals, despite the weaker resonant coupling strengths in ZnPc-MOF, suggesting their enhanced contribution to low-energy electronic behavior.

Although this work focuses on ZnPc-MOF, the conclusions drawn, such as the classification of band structures and the OTSRs, are also applicable to other square-lattice 2D materials with the same space group symmetry. Furthermore, the theoretical framework based on symmetry analysis and group representation theory can be extended to investigate the electronic and optical properties of 2D materials with different symmetries.

Noted that polarization-dependent OTSRs can give rise to valley polarization, and this phenomenon can be probed through valley-decomposed optical absorption spectra. However, since ZnPc-MOF exhibits a direct band gap at the $M$ point, it lacks valley degree of freedom. In contrast, consider another square-lattice 2D material that shares the same space group symmetry but, importantly, hosts valleys at the $X$ and $X^{\prime}$ points rather than at $M$ or $\Gamma$. In such a system, the valley-decomposed optical absorption spectra would clearly exhibit valley polarization.

The small band dispersion of the ZnPc-MOF monolayer brings the quasicrystalline electronic states closer to the band gap, but it also causes these states to coexist with extended crystalline states within the same energy window, thereby limiting their contribution to transport properties. For a 2D material with a square or hexagonal lattice, if its valleys are located at the $Q$ points, the intersections of the BZ boundaries of the two layers, one would expect the quasicrystalline electronic states to constitute the VBM and CBM of the vdW quasicrystal. This would significantly enhance their role in low-energy physics and transport phenomena.

\begin{acknowledgments}
This work was supported by National Natural Science Foundation of China (Grant No. 12204092), Science and Technology Development Plan Project of Jilin Province, China (Grant No. 20240101323JC), and the Fundamental Research Funds for the Central Universities (Grant No. 2412025QD011). The numerical calculations were performed on the supercomputing system of Northeast Normal University.
\end{acknowledgments}

\appendix

\section{Hamiltonian at k for systems with periodicity}
\label{app:Hk}
At wave vector $\bm{k}$ in the BZ, we define the Bloch functions of the system for sublattice $\mathcal{X}$ as
\begin{equation}
\varphi_{\bm{k},\mathcal{X}}(\bm{r}) = \langle \bm{r} | \bm{k}, \mathcal{X} \rangle = \frac{1}{\sqrt{M}} \sum_{\bm{\mathcal{R}}} e^{i\bm{k} \cdot (\bm{\mathcal{R}} + \bm{\tau}_{\mathcal{X}})} \; p_z(\bm{r} - \bm{\mathcal{R}} - \bm{\tau}_{\mathcal{X}}),
\end{equation}
where $M$ is the number of unit cells, $\bm{\mathcal{R}}$ denotes a lattice vector, $\bm{\tau}_{\mathcal{X}}$ is the position of sublattice $\mathcal{X}$ within the unit cell, and $p_z(\bm{r} - \bm{\mathcal{R}} - \bm{\tau}_{\mathcal{X}})$ represents the $p_z$ orbital centered at $\bm{\mathcal{R}} + \bm{\tau}_{\mathcal{X}}$. 

Using these Bloch functions as bases, the Hamiltonian in Eq.~\eqref{eq:H} becomes block-diagonal in $\bm{k}$-space and can be written as $H = \sum_{\bm{k}} H(\bm{k})$, with
\begin{equation}
H(\bm{k}) = \sum_{\mathcal{X}, \mathcal{X}^{\prime}} H_{\mathcal{X}\mathcal{X}^{\prime}}(\bm{k}) \; | \bm{k}, \mathcal{X} \rangle \langle \bm{k}, \mathcal{X}^{\prime} |,
\end{equation}
where the matrix elements are given by
\begin{equation}
H_{\mathcal{X}\mathcal{X}^{\prime}}(\bm{k}) = \sum_{\bm{\mathcal{R}}} t(\bm{\mathcal{R}} + \bm{\tau}_{\mathcal{X}^{\prime}} - \bm{\tau}_{\mathcal{X}}) \; e^{i\bm{k} \cdot (\bm{\mathcal{R}} + \bm{\tau}_{\mathcal{X}^{\prime}} - \bm{\tau}_{\mathcal{X}})}.
\end{equation}
Diagonalization of $H(\bm{k})$ yields the electronic band structures.

\section{Symmetry operation matrixes and irrep determination}
\label{app:irrep}
For each system considered, monolayer, AA- and AB-stacked, or twisted bilayer, we define the little group of a wavevector $\bm{k}$ as the set of all symmetry operations $\{R|\bm{u}\} \in \mathbb{G}_0$ (with $\bm{u} = 0$ for SSG systems and $\bm{u} = 0$ or $\bm{\tau}_0$ for non-SSG system) that satisfy $R\bm{k} = \bm{k} + \bm{G}$ for some reciprocal lattice vector $\bm{G}$. 

Using the Bloch functions employed in constructing the Hamiltonian $H(\bm{k})$ as shown in Appendix \ref{app:Hk}, the representation matrix of a symmetry operation $\{R|\bm{u}\}$ in the little group takes the form:
\begin{equation}
\begin{split}
D_{\mathcal{X}_1,\mathcal{X}_2}^{\bm{k}}(\{R|\bm{u}\}) &\equiv \langle \bm{k},\mathcal{X}_1 | P_{\{R|\bm{u}\}} | \bm{k},\mathcal{X}_2 \rangle \\
&= \zeta \, e^{-i\bm{k} \cdot \bm{u}} \, e^{i\bm{\mathcal{G}} \cdot \bm{\tau}_{\mathcal{X}_2^{\prime}}} \, \delta_{\mathcal{X}_1, \mathcal{X}_2^{\prime}},
\label{eq:D}
\end{split}
\end{equation}
where $R\bm{k} = \bm{k} + \bm{\mathcal{G}}$, $R\bm{\tau}_{\mathcal{X}_2} = \bm{\mathcal{R}} + \bm{\tau}_{\mathcal{X}_2^{\prime}}$, and $P_{\{R|\bm{u}\}}$ is the operator acting on the functions corresponding to $\{R|\bm{u}\}$. The factor $\zeta$ takes the value $+1$ or $-1$ depending on whether $\{R|\bm{u}\}$ preserves or swaps the layer index, respectively. The matrices in Eq.~\eqref{eq:D} form a linear (projective) representation of the little group at $\bm{k}$ for SSG (non-SSG) systems.

Since the representation matrix of each symmetry operation in the little group commutes with the Hamiltonian matrix $H(\bm{k})$, the eigenstates of $H(\bm{k})$ can be classified according to the irreducible representations (for SSG systems) or irreducible projective representations (for non-SSG system) of the little group at $\bm{k}$; both are hereafter abbreviated as \textit{irreps}. Within the framework of group representation theory, the irrep of an eigenstate can be identified using the projection operator:
\begin{equation}
P_{ir} = \frac{l_{ir}}{g} \sum_R [\chi_{ir}(R)]^* D(R),
\label{eq:proj_op}
\end{equation}
which satisfies $P_{ir} \Phi_{ir^{\prime}} = \Phi_{ir^{\prime}} \delta_{ir, ir^{\prime}}$. Here, $\Phi_{ir^{\prime}}$ denotes a state belonging to irrep $ir^{\prime}$, $l_{ir}$ is the dimension of irrep $ir$, $g$ is the order of the little group, $\chi_{ir}(R)$ is the character of symmetry operation $R$ in irrep $ir$, and $D(R)$ is the representation matrix defined in Eq.~\eqref{eq:D}. Character tables for the little groups at high-symmetry points and along high-symmetry lines are provided in Append \ref{app:tables}.

\section{Character tables and compatibility relationships}
\label{app:tables}
\counterwithin{table}{section}
\setcounter{table}{0}
\renewcommand{\thetable}{\Alph{section}\arabic{table}}

In Tables \ref{tab:irrep_monolayer}--\ref{tab:irrep_AB_SDDp}, we present the character tables of the little groups for wavevectors at high-symmetry points and along high-symmetry lines for monolayer, AA-stacked, AB-stacked, and commensurate twisted bilayer systems. Table \ref{tab:irrep_d4d} lists the character table of the point group $\mathbb{D}_{4d}$ for the ZnPc-MOF quasicrystal. In Table \ref{tab:compi_relationship}, we provide the CRs connecting high-symmetry lines with their two endpoints (high-symmetry points) for all four commensurate systems.

Tables \ref{tab:irrep_monolayer}, \ref{tab:irrep_AA}, and \ref{tab:irrep_Twistedbilayer} are for the monolayer, AA-stacked bilayer, and twisted bilayer, respectively, and they follow the same format. To illustrate the structure of these character tables, we take the example of $\bm{k}=\Gamma$ for the monolayer. Table \ref{tab:irrep_monolayer} indicates that $\Gamma$ and $M$ share the same little group and therefore the same character table. This little group is isomorphic to the point group $\mathbb{C}_{4v}$, which comprises eight symmetry operations: $E$, $C_{4z}$, $C_{4z}^3$, $C_{2z}$, $\sigma_{xz}$, $\sigma_{yz}$, $\sigma_{(xy)z}$, and $\sigma_{(\overline{x}y)z}$. $\mathbb{C}_{4v}$ has four 1D irreps $A_1$, $A_2$, $B_1$, $B_2$, and one 2D irrep $E$. The essential information for any irrep is the characters of all symmetry operations (or group elements in group-theoretic language), which are used to construct projection operators and to label electronic states. For all 1D irreps, the character of the identity operation $E$ is always 1, while for the 2D irrep it is always 2. Operations belonging to the same class share identical characters; thus $C_{4z}$ and $C_{4z}^3$ have the same character, as do $\sigma_{(xy)z}$ and $\sigma_{(\overline{x}y)z}$, and also $\sigma_{xz}$ and $\sigma_{yz}$. The characters of $\sigma_{xz}$ and $\sigma_{yz}$ are listed separately in the table, because the character tables for all high-symmetry points and lines are combined into one table, and they belong to different classes for the irreps $B_1$ and $B_2$ at $\bm{k}=X$ and $X^{\prime}$.

For the AB-stacked bilayer, the character tables are provided separately for different sets of wavevectors: those at $\Gamma$ and $M$ are given in Table \ref{tab:irrep_AB_GM}; those at $X$, $X^{\prime}$, $Y$, and $Y^{\prime}$ in Table \ref{tab:irrep_AB_XY}; and those along $\Delta$, $\Delta^{\prime}$, and $\Sigma$ in Table \ref{tab:irrep_AB_SDDp}. In each subtable of these tables, the characters of all group elements belonging to the same conjugacy class are grouped together. It is worth noting that the little groups for $M$, $X$, $X^{\prime}$, $Y$, and $Y^{\prime}$ possess only 2D irreps; consequently, bands at these $\bm{k}$ points are always twofold degenerate.

Given a band along the high-symmetry line with endpoints at high-symmetry points, the CRs in Table \ref{tab:compi_relationship} offer a way to predict and understand the relation between the irreps on the line and at the endpoints. As an example, consider the CR for ($\Gamma$, $\Delta$, $X$) of the monolayer. This relation describes bands along the $\Delta$ line, with endpoints at $\Gamma$ and $X$ [see the bands in Fig. \ref{fig:band_irrep}(a)]. It states that if a band along $\Delta$ has irrep $A^{\prime}$, then the irrep at $\Gamma$ must be either $A_1$ or $B_1$ (assuming it is non-degenerate), and the irrep at $X$ must be either $A_1$ or $B_2$. Conversely, if the irrep along $\Delta$ is $A^{\prime\prime}$, the irrep at $\Gamma$ must be $A_2$ or $B_2$ (in the non-degenerate case), and the irrep at $X$ must be $A_2$ or $B_1$. Moreover, if there is a twofold degenerate energy level with the irrep $E$ at $\Gamma$, it splits into two non-degenerate levels along $\Delta$, one with $A^{\prime}$ symmetry and the other with $A^{\prime\prime}$ symmetry.

\begin{table}[htb]
\caption{Character tables of little groups for high-symmetry points and high-symmetry lines for monolayer.}
\label{tab:irrep_monolayer}
\centering
\begin{tabular*}{\linewidth}{@{\extracolsep{\fill}}lll|cccccc@{}}
\toprule
$\mathbb{C}_s$ & $\mathbb{C}_{2v}$ & $\mathbb{C}_{4v}$ & \multicolumn{6}{c}{Monolayer} \\
\midrule
$Y|\Delta^{\prime}$      &   &              & $E$ &                   &          & $\sigma_{xz}$ & &\\
$\Delta|Y^{\prime}$ &   &              & $E$ &                   &          &            & $\sigma_{yz}$ & \\
$\Sigma$ &   &              & $E$ &                   &          &               & &$\sigma_{(xy)z}$  \\
         &$X|X^{\prime}$&              & $E$ &                   & $C_{2z}$ & $\sigma_{xz}$&$\sigma_{yz}$      & \\
         &   & $M|\Gamma$ & $E$ & \makecell{$C_{4z}$\\$C_{4z}^3$} & $C_{2z}$ & $\sigma_{xz}$  &$\sigma_{yz}$& \makecell{$\sigma_{(xy)z}$\\$\sigma_{(\overline{x}y)z}$} \\
\midrule
$A^{\prime}$       &$A_1$ & $A_1$ & 1 & 1 & 1 & 1 & 1 & 1 \\
$A^{\prime\prime}$ &$A_2$ & $A_2$ & 1 & 1 & 1 &-1 &-1 &-1  \\
                   &      & $B_1$ & 1 &-1 & 1 & 1 & 1 &-1  \\
                   &      & $B_2$ & 1 &-1 & 1 &-1 &-1 & 1  \\
                   &      & $E$   & 2 & 0 &-2 & 0 & 0 & 0  \\
                   &$B_1$ &       & 1 &   &-1 & 1 &-1 & \\
                   &$B_2$ &       & 1 &   &-1 &-1 & 1 & \\
\bottomrule
\end{tabular*}
\end{table}

\begin{table*}[htbp]
\caption{Character tables of little groups for high-symmetry points and lines for AA-stacked bilayer.}
\label{tab:irrep_AA}
\centering
\begin{tabular*}{\linewidth}{@{\extracolsep{\fill}}lll|ccccccccccccc@{}}
\toprule
$\mathbb{C}_{2v}$ & $\mathbb{D}_{2h}$ & $\mathbb{D}_{4h}$ & \multicolumn{13}{c}{AA-stacked bilayer} \\
\midrule
$Y|\Delta^{\prime}$ & & & $E$ & & & $C_{2x}$ && & & &$\sigma_{xy}$ & $\sigma_{xz}$ & & & \\
$\Delta|Y^{\prime}$ & & & $E$ & & & &$C_{2y}$ & & & & $\sigma_{xy}$ & & $\sigma_{yz}$ & & \\
$\Sigma$ & & & $E$ & & & &&$C_{2xy}$ & & & $\sigma_{xy}$ & & & $\sigma_{(xy)z}$ & \\
 & $X|X^{\prime}$ & & $E$ & & $C_{2z}$ & $C_{2x}$&$C_{2y}$ & & $i$ & & $\sigma_{xy}$ & $\sigma_{xz}$&$\sigma_{yz}$ & & \\
 & & $\Gamma|M$ & $E$ & \thead{$C_{4z}$ \\$C_{4z}^3$}& $C_{2z}$ & $C_{2x}$ & $C_{2y}$& \thead{$C_{2xy}$\\$C_{2\overline{x}y}$ }& $i$ & \thead{$S_{4z}$\\$S_{4z}^3$} & $\sigma_{xy}$ & $\sigma_{xz}$ &$\sigma_{yz}$& \thead{$\sigma_{(xy)z}$\\$\sigma_{(\overline{x}y)z}$} & \\
\midrule
$A_1$ & $A_g$     & $A_{1g}$ & 1 &  1 &  1 &  1 &  1 &  1 &  1 &  1 &  1 &  1 &  1 &  1 \\
$B_1$ & $B_{1g}$  & $A_{2g}$ & 1 &  1 &  1 & -1 & -1 & -1 &  1 &  1 &  1 & -1 & -1 & -1 \\
      &           & $B_{1g}$ & 1 & -1 &  1 &  1 &  1 & -1 &  1 & -1 &  1 &  1 &  1 & -1 \\
      &           & $B_{2g}$ & 1 & -1 &  1 & -1 & -1 &  1 &  1 & -1 &  1 & -1 & -1 &  1 \\
      &           & $E_g$    & 2 &  0 & -2 &  0 &  0 &  0 &  2 &  0 & -2 &  0 &  0 &  0 \\
$A_2$ & $A_u$     & $A_{1u}$ & 1 &  1 &  1 &  1 &  1 &  1 & -1 & -1 & -1 & -1 & -1 & -1 \\
$B_2$ & $B_{1u}$  & $A_{2u}$ & 1 &  1 &  1 & -1 & -1 & -1 & -1 & -1 & -1 &  1 &  1 &  1 \\
      &           & $B_{1u}$ & 1 & -1 &  1 &  1 &  1 & -1 & -1 &  1 & -1 & -1 & -1 &  1 \\
      &           & $B_{2u}$ & 1 & -1 &  1 & -1 & -1 &  1 & -1 &  1 & -1 &  1 &  1 & -1 \\
      &           & $E_{u}$  & 2 &  0 & -2 &  0 &  0 &  0 & -2 &  0 &  2 &  0 &  0 &  0 \\
      & $B_{2g}$  &          & 1 &    & -1 & -1 &  1 &    &  1 &    & -1 &  1 & -1 &\\
      & $B_{3g}$  &          & 1 &    & -1 &  1 & -1 &    &  1 &    & -1 & -1 &  1 &\\
      & $B_{2u}$  &          & 1 &    & -1 & -1 &  1 &    & -1 &    &  1 & -1 &  1 &\\
      & $B_{3u}$  &          & 1 &    & -1 &  1 & -1 &    & -1 &    &  1 &  1 & -1 &\\
\bottomrule
\end{tabular*}
\end{table*}

\begin{table}[htbp]
\caption{Character tables of little groups for high-symmetry points and lines for twisted bilayer.}
\label{tab:irrep_Twistedbilayer}
\centering
\begin{tabular*}{\linewidth}{@{\extracolsep{\fill}}lll|cccccc@{}}
\toprule
$\mathbb{C}_2$ & $\mathbb{D}_2$ & $\mathbb{D}_4$ &   \multicolumn{6}{c}{Twisted bilayer}\\
\midrule
$Y|\Delta^{\prime}$            &                &                & $E$ &               &         &$C_{2x}$ &      &\\
$\Delta|Y^{\prime}$       &                &                & $E$ &               &         &          &$C_{2y}$       &\\
$\Sigma$       &                &                & $E$ &               &         &           &    &$C_{2xy}$\\
               &   $X|X^{\prime}$          &                & $E$ &               &$C_{2z}$&$C_{2x}$&$C_{2y}$                    \\
               &                &   $\Gamma|M$   & $E$ &\thead{$C_{4z}$\\$C_{4z}^3$}&$C_{2z}$&$C_{2x}$&$C_{2y}$&\thead{$C_{2xy}$\\$C_{2\overline{x}y}$}\\
\midrule
     $A$       &   $A$          &   $A_1$        & 1 &  1      &    1   &       1 & 1       &   1 \\
     $B$       &   $B_1$        &   $A_2$        & 1 &  1      &    1   &      -1 &-1      &  -1 \\
               &                &   $B_1$        & 1 & -1      &    1   &       1 & 1      &  -1 \\
               &                &   $B_2$        & 1 & -1      &    1   &      -1 &-1      &   1 \\
               &                &   $E$          & 2 &  0      &   -2   &       0 & 0      &   0 \\
               &   $B_2$        &                & 1 &         &   -1   &      -1 & 1     &     \\
               &   $B_3$        &                & 1 &         &   -1   &       1 &-1     &     \\
\bottomrule
\end{tabular*}
\end{table}

\begin{table*}[!htbp]
\centering
\caption{Character table of little groups at $\Gamma$ and $M$ for AB-stacked bilaye.}
\label{tab:irrep_AB_GM}
\begin{tabular*}{\linewidth}{@{\extracolsep{\fill}}lcccccccccc@{}}
\toprule
$\Gamma$ & $E$ & \makecell{$C_{4z}$\\$C_{4z}^3$} & $C_{2z}$ & \makecell{$\{C_{2x}|\bm{\tau}_0\}$\\$\{C_{2y}|\bm{\tau}_0\}$} & \makecell{$\{C_{2xy}|\bm{\tau}_0\}$\\$\{C_{2\overline{x}y}|\bm{\tau}_0\}$} & $\{i|\bm{\tau}_0\}$ & \makecell{$\{S_{4z}|\bm{\tau}_0\}$\\$\{S_{4z}^3|\bm{\tau}_0\}$} & $\{\sigma_{xy}|\bm{\tau}_0\}$ & \makecell{$\sigma_{yz}$\\$\sigma_{xz}$} & \makecell{$\sigma_{(\overline{x}y)z}$\\$\sigma_{(xy)z}$} \\
\midrule
$\Gamma_1^+$ & 1 & 1 & 1 & 1 & 1 &  1 &  1 &  1 & 1 & 1 \\
$\Gamma_1^-$ & 1 & 1 & 1 & 1 & 1 & -1 & -1 & -1 & -1 & -1 \\
$\Gamma_2^+$ & 1 &-1 & 1 & 1 &-1 &  1 & -1 &  1 & 1 & -1 \\
$\Gamma_2^-$ & 1 &-1 & 1 & 1 &-1 & -1 &  1 & -1 & -1 & 1 \\
$\Gamma_3^+$ & 1 & 1 & 1 &-1 &-1 &  1 &  1 &  1 & -1 & -1 \\
$\Gamma_3^-$ & 1 & 1 & 1 &-1 &-1 & -1 & -1 & -1 & 1 & 1 \\
$\Gamma_4^+$ & 1 &-1 & 1 &-1 & 1 &  1 & -1 &  1 & -1 & 1 \\
$\Gamma_4^-$ & 1 &-1 & 1 &-1 & 1 & -1 &  1 & -1 & 1 & -1 \\
$\Gamma_5^+$ & 2 & 0 &-2 & 0 & 0 &  2 &  0 & -2 & 0 & 0 \\
$\Gamma_5^-$ & 2 & 0 &-2 & 0 & 0 & -2 &  0 &  2 & 0 & 0 \\
\bottomrule
\end{tabular*}
\begin{tabular*}{\linewidth}{@{\extracolsep{\fill}}lccccccccccc@{}}
\toprule
$M$ & $E$ & \makecell{$C_{4z}$\\$C_{4z}^3$} & $C_{2z}$ & \makecell{$\{C_{2x}|\bm{\tau}_0\}$\\$\{C_{2y}|\bm{\tau}_0\}$} & $\{C_{2xy}|\bm{\tau}_0\}$ & $\{C_{2\overline{x}y}|\bm{\tau}_0\}$ & $\{i|\bm{\tau}_0\}$ & \makecell{$\{S_{4z}|\bm{\tau}_0\}$\\$\{S_{4z}^3|\bm{\tau}_0\}$} & $\{\sigma_{xy}|\bm{\tau}_0\}$ & \makecell{$\sigma_{yz}$\\$\sigma_{xz}$} & \makecell{$\sigma_{(\overline{x}y)z}$\\$\sigma_{(xy)z}$} \\
\midrule
$M_1$ & 2 & 0 & -2 & 0 &  2 & -2 &  0 &  0 & 0 & 0 & 0 \\
$M_2$ & 2 & 0 & -2 & 0 & -2 &  2 &  0 &  0 & 0 & 0 & 0 \\
$M_3$ & 2 & 0 &  2 & 0 &  0 &  0 &  0 &  0 & 0 & 0 & 2 \\
$M_4$ & 2 & 0 &  2 & 0 &  0 &  0 &  0 &  0 & 0 & 0 & -2 \\
\bottomrule
\end{tabular*}
\end{table*}

\begin{table}[!htbp]
\centering
\caption{Character table of little groups at high-symmetry points $X$ and $X^{\prime}$ and on high-symmetry lines $Y$ and $Y^{\prime}$ for AB-stacked bilayer.}
\label{tab:irrep_AB_XY}
\centering
\begin{tabular*}{\linewidth}{@{\extracolsep{\fill}}lcccccccc@{}}
\toprule
$X$ & $E$ & $C_{2z}$ & $\{C_{2x}|\bm{\tau}_0\}$ & $\{C_{2y}|\bm{\tau}_0\}$ & $\{i|\bm{\tau}_0\}$ & $\{\sigma_{xy}|\bm{\tau}_0\}$ & $\sigma_{yz}$ & $\sigma_{xz}$ \\
\midrule
$X_1$ & 2 & 0 & 0 & 0 & 0 & 0 &  2 & 0 \\
$X_2$ & 2 & 0 & 0 & 0 & 0 & 0 & -2 & 0 \\
\bottomrule
$X^{\prime}$ & $E$ & $C_{2z}$ & $\{C_{2x}|\bm{\tau}_0\}$ & $\{C_{2y}|\bm{\tau}_0\}$ & $\{i|\bm{\tau}_0\}$ & $\{\sigma_{xy}|\bm{\tau}_0\}$ & $\sigma_{yz}$ & $\sigma_{xz}$ \\
\midrule
$X_1^{\prime}$ & 2 & 0 & 0 & 0 & 0 & 0 &  0 & 2 \\
$X_2^{\prime}$ & 2 & 0 & 0 & 0 & 0 & 0 &  0 & -2 \\
\bottomrule
\end{tabular*}
\begin{tabular*}{\linewidth}{@{\extracolsep{\fill}}lcccc@{}}
\toprule
$Y$ & $E$ & $\{C_{2x}|\bm{\tau}_0\}$ & $\{\sigma_{xy}|\bm{\tau}_0\}$ & $\sigma_{xz}$ \\
\midrule
$Y_1$ & 2 & 0 & 0 & 0 \\
\bottomrule
$Y^{\prime}$ & $E$ & $\{C_{2y}|\bm{\tau}_0\}$ & $\{\sigma_{xy}|\bm{\tau}_0\}$ & $\sigma_{yz}$ \\
\midrule
$Y_1^{\prime}$ & 2 & 0 & 0 & 0 \\
\bottomrule
\end{tabular*}
\end{table}

\begin{table}[!htbp]
\centering
\small
\caption{Character table of little groups on symmetry lines $\Sigma$, $\Delta$, and $\Delta^{\prime}$ for AB-stacked bilayer, for which we take wavevector at $\bm{k}=u(\bm{b}_1+\bm{b}_2)$, $\bm{k}=u\bm{b}_2$, and $\bm{k}=u\bm{b}_1$, respectively.}
\label{tab:irrep_AB_SDDp}
\centering
\begin{tabular*}{\linewidth}{@{\extracolsep{\fill}}lcccc@{}}
\toprule
$\Sigma$ & $E$  & $\{C_{2xy}|\bm{\tau}_0\}$ & $\{\sigma_{xy}|\bm{\tau}_0\}$&$\sigma_{(xy)z}$ \\
\midrule
$\Sigma_1$ & 1 & $e^{i2\pi u}$ & $e^{i2\pi u}$ &1\\
$\Sigma_2$ & 1 & $e^{i\pi(1+2u)}$ & $e^{i2\pi u}$ &-1\\
$\Sigma_3$ & 1 & $e^{i\pi(1+2u)}$ & $e^{i\pi(1+2u)}$ &1\\
$\Sigma_4$ & 1 & $e^{i2\pi u}$ & $e^{i\pi(1+2u)}$ &-1 \\
\bottomrule
$\Delta$ & $E$ & $\{C_{2y}|\bm{\tau}_0\}$ & $\{\sigma_{xy}|\bm{\tau}_0\}$ & $\sigma_{yz}$ \\
\midrule
$\Delta_1$ & 1 & $e^{i2\pi u}$ & $e^{i2\pi u}$ & 1 \\
$\Delta_2$ & 1 & $e^{i2\pi u}$ & $e^{i\pi (1+u)}$ & -1 \\
$\Delta_3$ & 1 & $e^{i\pi (1+u)}$ & $e^{i\pi (1+u)}$ & 1 \\
$\Delta_4$ & 1 & $e^{i\pi (1+u)}$ & $e^{i\pi u}$ & -1 \\
\bottomrule
$\Delta^{\prime}$ & $E$ & $\{C_{2x}|\bm{\tau}_0\}$ & $\{\sigma_{xy}|\bm{\tau}_0\}$ & $\sigma_{xz}$ \\
\midrule
$\Delta_1^{\prime}$ & 1 & $e^{i2\pi u}$ & $e^{i2\pi u}$ & 1 \\
$\Delta_2^{\prime}$ & 1 & $e^{i2\pi u}$ & $e^{i\pi (1+u)}$ & -1 \\
$\Delta_3^{\prime}$ & 1 & $e^{i\pi (1+u)}$ & $e^{i\pi (1+u)}$ & 1 \\
$\Delta_4^{\prime}$ & 1 & $e^{i\pi (1+u)}$ & $e^{i\pi u}$ & -1 \\
\bottomrule
\end{tabular*}
\end{table}

\begin{table}[!htbp]
\caption{Character table of $\mathbb{D}_{4d}$ point group for 45$^{\circ}$ twisted bilayer. The four 2-fold rotations with axes inside the $xy$-plane have the same characters and given in $4C_{2}^{\prime}$ together. The four mirror reflections with mirror containing $z$ axis have the same character, and their characters are given in $4\sigma_d$ together.}
{\begin{tabular*}{\linewidth}{@{\extracolsep{\fill}}ccccccccc@{}}
\toprule
$\mathbb{D}_{4d}$ & $E$ & \thead{$S_{8z}$\\$S_{8z}^7$} &\thead{ $C_{4z}$ \\$C_{4z}^3$} &\thead{ $S_{8z}^{3}$\\$S_{8z}^5$} & $C_{2z}$ & $4C_{2}^{\prime}$ & $4\sigma_d$ \\
\midrule
$A_1$ &  1  &      1           &      1            &       1          &    1     &       1          &   1   \\
$A_2$ &  1  &      1           &      1            &       1          &    1     &      -1          &  -1   \\
$B_1$ &  1  &     -1           &      1            &      -1          &    1     &       1          &  -1   \\
$B_2$ &  1  &     -1           &      1            &      -1          &    1     &      -1          &   1   \\
$E_1$ &  2  &   $\sqrt{2}$     &      0            &   $-\sqrt{2}$    &   -2     &       0          &   0   \\
$E_2$ &  2  &      0           &     -2            &       0          &    2     &       0          &   0   \\
$E_3$ &  2  &   $-\sqrt{2}$    &      0            &    $\sqrt{2}$    &   -2     &       0          &   0   \\
\bottomrule
\end{tabular*}}
\label{tab:irrep_d4d}
\end{table}

\begin{table*}[htbp]
\caption{CRs between high-symmetry lines and high-symmetry points. The CR for the pairs ($\Gamma,\Delta,X$) and ($\Gamma,\Delta^{\prime},X^{\prime}$) is collectively denoted as ($\Gamma, \Delta^{(\prime)}, X^{(\prime)}$). Similarly, ($M,Y,X$) and ($M,Y^{\prime},X^{\prime}$) are labeled as ($M,Y^{(\prime)},X^{(\prime)}$). As an example, consider the CR ($\Gamma, \Delta^{(\prime)}, X^{(\prime)}$) for monolayer. At $\Delta$, the $A^{\prime}$ state is compatible with the $A_1$ and $B_1$ states at $\Gamma$, and with the $A_1$ and $B_2$ states at $X$. At $\Delta^{\prime}$, the $A^{\prime}$ state is compatible with the $A_1$ and $B_1$ states at $\Gamma$, and with the $A_1$ and $B_1$ states at $X^{\prime}$.}
{\begin{tabular*}{\linewidth}{@{\extracolsep{\fill}}c|ccc|ccc|ccc@{}}
\toprule
&$\Gamma$ & $\Sigma$ & $M$ & $\Gamma$ &$\Delta^{(\prime)}$ & $X^{(\prime)}$ & $M$ & $Y^{(\prime)}$ & $X^{(\prime)}$\\
\midrule
                      &$A_1,B_2$ & $A^{\prime}$ & $A_1,B_2$ &$A_1,B_1$ &$A^{\prime}$ &$A_1,B_2(B_1)$& $A_1,B_1$& $A^{\prime}$& $A_1,B_1(B_2)$ \\
$Monolayer$&$A_2,B_1$ & $A^{\prime\prime}$ & $A_2,B_1$ &$A_2,B_2$ &$A^{\prime\prime}$ &$A_2,B_1(B_2)$ & $A_2,B_2$ & $A^{\prime\prime}$ & $A_2,B_2(B_1)$ \\
                      &$E$ & $A^{\prime}\oplus A^{\prime\prime}$ & $E$ & $E$ & $A^{\prime}\oplus A^{\prime\prime}$ & & $E$& $A^{\prime}\oplus A^{\prime\prime}$ & \\
\midrule
    &$A_{1g},B_{2g}$ & $A_1$ & $A_{1g},B_{2g}$ &$A_{1g},B_{1g}$ &$A_1$ &$A_g,B_{2u}(B_{3u})$ & $A_{1g},B_{1g} $& $A_1$& $A_g,B_{3u}(B_{2u})$ \\
    &$A_{2g},B_{1g}$ & $B_1$ & $A_{2g},B_{1g}$ &$A_{2g},B_{2g}$ &$B_1$ &$B_{1g},B_{3u}(B_{2u})$ & $A_{2g},B_{2g}$ & $B_1$ & $B_{1g},B_{2u}(B_{3u})$  \\
$AA$&$A_{1u},B_{2u}$ & $A_2$ & $A_{1u},B_{2u}$ &$A_{1u},B_{1u}$ & $A_2$&$A_u,B_{2g}(B_{3g})$ & $A_{1u},B_{1u}$ & $A_2$ & $A_u,B_{3g}(B_{2g})$\\
    &$A_{2u},B_{1u}$ & $B_2$ & $A_{2u},B_{1u}$ &$A_{2u},B_{2u}$ & $B_2$&$B_{1u},B_{3g}(B_{2g})$ &$A_{2u},B_{2u}$ & $B_2$ & $B_{1u},B_{2g}(B_{3g})$\\
    &$E_g$ & $A_2\oplus B_2$ & $E_g$ &$E_g$ & $A_2\oplus B_2$&&$E_g$ & $A_2\oplus B_2$ &\\
    &$E_u$ & $A_1\oplus B_1$ & $E_u$ &$E_u$ & $A_1\oplus B_1$&&$E_u$ & $A_1\oplus B_1$ &\\
\midrule
    &$\Gamma_1^+,\Gamma_4^+$ & $\Sigma_1$ & -- &$\Gamma_1^+,\Gamma_2^+$ &$\Delta_1^{(\prime)}$ &-- & $M_{1,2,3,4}$& $Y_1^{(\prime)}$& $X_1^{(\prime)},X_2^{(\prime)}$ \\
&-- & $\Sigma_1\oplus \Sigma_4$ & $M_1$ &-- & $\Delta_1^{(\prime)} \oplus \Delta_3^{(\prime)}$ & $X_1^{(\prime)}$ & & &  \\
    &$\Gamma_2^+,\Gamma_3^+$ & $\Sigma_2$ & -- &$\Gamma_1^-,\Gamma_2^-$ &$\Delta_2^{(\prime)}$ &-- &  &  &   \\
    &-- & $\Sigma_2\oplus \Sigma_3$ & $M_2$ &-- & $\Delta_2^{(\prime)} \oplus \Delta_4^{(\prime)}$ &$X_2^{(\prime)}$ &  & &   \\
$AB$&$\Gamma_2^-,\Gamma_3^-$ & $\Sigma_3$ & -- &$\Gamma_3^-,\Gamma_4^-$ &$\Delta_3^{(\prime)}$ & -- &  &  & \\
    &-- & $\Sigma_1\oplus \Sigma_3$ & $M_3$ &$\Gamma_3^+,\Gamma_4^+$ &$\Delta_4^{(\prime)}$ & -- & & & \\
    &$\Gamma_1^-,\Gamma_4^-$ & $\Sigma_4$ & -- &$\Gamma_5^+$ & $\Delta_2^{(\prime)}\oplus \Delta_3^{(\prime)}$ & -- & & & \\
    &-- & $\Sigma_2\oplus \Sigma_4$ & $M_4$ &$\Gamma_5^-$ &$\Delta_1^{(\prime)}\oplus \Delta_4^{(\prime)}$ & --& & & \\
    &$\Gamma_5^+$ & $\Sigma_3\oplus \Sigma_4$ & -- & & && & &\\
    &$\Gamma_5^-$ & $\Sigma_1\oplus \Sigma_2$ & -- & & && & &\\
\midrule
\multirow{3}{*}{\thead{$Twisted$\\ $bilayer$}} & $A_1,B_2$ & $A$ & $A_1,B_2$ & $A_1,B_1$ & $A$ & $A,B_2(B_3)$& $A$ &$A$&$A,B_3(B_2)$\\
         &  $A_2,B_1$ & $B$ & $A_2,B_1$ & $A_2,B_2$ & $B$ & $B_1,B_3(B_2)$& $A_1,B_1$ &$B$&$B_1,B_2(B_3)$\\
          &         $E$ & $A\oplus B$ & $E$     & $E$ & $A\oplus B$ &        & $E$ &$A\oplus B$&\\
\bottomrule
\end{tabular*}}
\label{tab:compi_relationship}
\end{table*}


%

\end{document}